\documentclass{emulateapj}

\shorttitle{Global structure of isothermal X-ray emission along the Fermi bubbles}  
\shortauthors{Kataoka et al.}
\usepackage{times}

\def\F{{\it Fermi}-LAT }
\def\SZ{{\it Suzaku }}
\def\SW{{\it Swift }}
\def\deg{\hbox{$^\circ$}}
\def\arcsec{\hbox{$^{\prime\prime}$}}

\begin{document}

\title{Global Structure of Isothermal Diffuse X-ray Emission\\
along the Fermi Bubbles}

\author{
J. Kataoka\altaffilmark{1,\,2}, M. Tahara\altaffilmark{1}, T. Totani\altaffilmark{3}, Y. Sofue\altaffilmark{4}, Y. Inoue\altaffilmark{5}, S. Nakashima\altaffilmark{5}, C. C. Cheung\altaffilmark{6}}
\altaffiltext{1}{Research Institute for Science and Engineering, Waseda University, 3-4-1, Okubo, Shinjuku, Tokyo 169-8555, Japan}
\altaffiltext{2}{email: \texttt{kataoka.jun@waseda.jp}}
\altaffiltext{3}{Department of Astronomy, The University of Tokyo, Bunkyo-ku, Tokyo 113-0033, Japan}
\altaffiltext{4}{Institute of Astronomy, The University of Tokyo, Mitaka, Tokyo 181-0015, Japan}
\altaffiltext{5}{Institute of Space and Astronautical Science, JAXA, 3-1-1 Yoshinodai, Chuo-ku, Sagamihara, Kanagawa 252-5210, Japan}
\altaffiltext{6}{Space Science Division, Naval Research Laboratory, Washington, DC 20375, USA}

\begin{abstract}
In our previous works (Kataoka  et al. 2013, Tahara et al. 2015), 
we found absorbed thermal X-ray plasma with 
$kT$ $\simeq$ 0.3 keV observed ubiquitously near the edges 
of the Fermi bubbles and interpreted this emission as 
weakly shock-heated Galactic 
halo (GH) gas. Here we present a systematic and uniform analysis 
of archival \textit{Suzaku} (29 pointings; 6 newly presented) 
and \textit{Swift} (68 pointings; 49 newly presented) data
within Galactic longitudes $|$$l$$|$ $<$ 20\deg\ and latitude 
5\deg $\lesssim$ $|$$b$$|$ 
$<$ 60\deg, covering the  whole extent of the Fermi bubbles.
We show that the plasma temperature is constant at  
$kT$ $\simeq$ 0.30$\pm$0.07 keV, while the emission measure 
(EM) varies by an order of magnitude, increasing 
toward the Galactic center 
(i.e., low $|b|$) with enhancements at the north polar spur (NPS), 
SE-claw and NW-clump features. Moreover, the EM distribution 
of $kT$ $\simeq$ 0.30 keV plasma is highly asymmetric in the northern 
and southern bubbles. Although the association of the X-ray emission 
with the bubbles is not conclusive, we compare 
the observed EM properties with simple models assuming 
(i) a filled halo without bubbles, whose gas density follows a hydrostatic 
isothermal model (King profile) 
and (ii) a bubble-in-halo in which two identical bubbles expand 
into the  halo forming thick shells of swept halo gas. 
We argue that the EM profile in the north ($b$ $>$ 0\deg) favors 
(ii), whereas that of the south ($b$ $<$ 0\deg) is rather 
close to (i), but weak excess signature is clearly detected 
also in the south like NPS (South Polar Spur; SPS). 
Such an asymmetry, if due to the bubbles,   
cannot be fully understood only 
by the inclination of bubbles' axis against the Galactic disk normal, 
thus suggesting asymmetric outflow due to 
different environmental/initial condition. 
\end{abstract}


\section{Introduction}

The ``Fermi bubbles'' are giant gamma-ray structures extending above and 
below the Galactic Center (GC) for about 8 kpc
(Dobler et al. 2010; Su et al. 2010; Ackermann et al. 2014). 
The gamma-ray emission of the bubbles is 
spatially correlated with the so-called ``WMAP haze'', which is characterized 
by a spherical morphology with radius $\sim$ 4~kpc centered at the GC, 
and was recently confirmed by $Planck$ observations (Planck Collaboration 
2013). Moreover, the recently discovered giant linearly-polarized radio lobes 
emanating from the GC also show a close correspondence to the Fermi bubbles 
(Carretti et al. 2013). It has thus been argued that the bubbles were created by 
some large episode of energy injection in the GC, such as an
AGN-like outburst (e.g., Guo et al. 2012; Yang et al. 2012) or from nuclear 
starburst activity (e.g., Lacki 2014) in the past with an energy release 
of 10$^{55-56}$ erg over 10 Myr ago 
(Su et al. 2010; Crocker \& Aharonian 2011; Carreti et al. 2013). 

Interestingly, the idea of a nuclear outburst which happened in the 
GC was first proposed over 40 years ago prior to the discovery of 
the Fermi bubbles (e.g., Sofue 1977; 1984; 1994; 2000; Bland-Hawthorn 
\& Cohen 2003). Relatedly, a number of observations in X-rays have been
discussed in the literature as evidence that the GC has experienced 
multiple epochs of enhanced source activity, including the $Fe$-K$_{\alpha}$ 
echo from molecular clouds (e.g., Koyama et al. 1996; Ryu et al. 2013)
and the presence of an 
over-ionized clump with a jet-like structure (Nakashima et al. 2013).
Particularly noteworthy is the giant Galactic feature called the North Polar Spur 
(NPS) that is seen both in X-ray and radio maps and believed to be a part of the radio Loop-I structure.
Sofue (2000) interpreted the NPS as a result of a large-scale outflow 
from the GC with a total energy of $\sim$10$^{55-56}$ erg within a 
timescale of $\sim 10$ Myr, exactly consistent with the values
discussed to create the Fermi bubbles above. In this context, 
Totani (2006) has shown that various other observational properties like
the 511 keV line emission (e.g., Weidenspointner et al. 2008) 
in the GC can also be naturally explained in the framework of a 
radiatively inefficient accretion flow (RIAF), if the outflow energy 
expected is 10$^{56}$ erg or 3$\times$10$^{41}$ erg s$^{-1}$.

\begin{figure*}[t]
\begin{center}
\includegraphics[angle=0,scale=0.7]{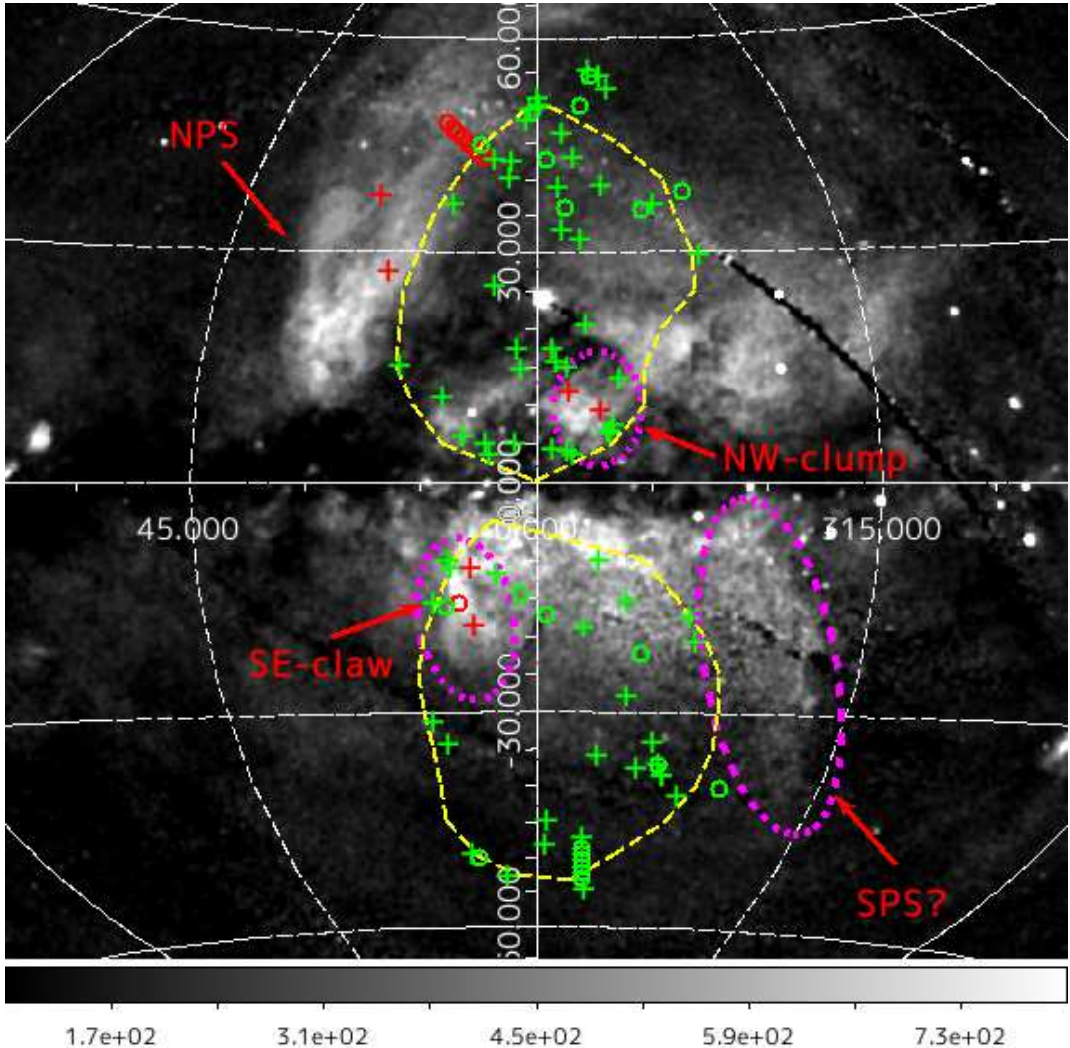}
\caption{Positions in Galactic coordinates of the 29 \textit{Suzaku} (circles) and 68 \textit{Swift} (crosses) X-ray data field of views 
systematically analyzed in this paper overlaid on a ROSAT 0.75 keV image (grayscale). 
The pointings within the NPS, SE-claw (an arc-shaped X-ray spur;  
dashed magenta) and NW-clump (an X-ray clump; dashed magenta), 
are shown in red, and all others are in green. Yellow dashed lines 
indicate the boundary of the Fermi bubbles, as suggested in Su et al. (2010). 
\label{fig:ptng}}
\end{center}
\end{figure*}

Assuming that the NPS and other prominent X-ray enhancements in the 
vicinity of the Fermi bubbles are all related in origin, we started 
a project consisting of X-ray observations along the 
edge regions of the Fermi bubbles since 2012, together with a systematic 
analysis of archival data provided by \SZ and \SW over the past 10 years. 
Kataoka et al. (2013; Paper-I) first carried out 14 \SZ X-ray observations 
positioned across the north-east and the  southern-most edges of the 
Fermi bubbles with a total requested exposure of 280 ksec. 
They found that the detected diffuse X-ray emission is reproduced 
by a three-component plasma model including unabsorbed thermal 
emission of the Local Bubble (LB: $kT$ $\simeq$ 0.1~keV), absorbed thermal 
emission related to the NPS and/or Galactic halo (GH: $kT$ $\simeq$ 0.3~keV), 
and a power-law component reproducing the cosmic X-ray background. 

This finding was confirmed by Tahara et al. (2015; Paper-II) who observed 
two other prominent X-ray structures, the North-cap (N-cap) and south-east claw 
(SE-claw) seen in the ROSAT 0.75 keV image 
(Snowden et al. 1995) and/or  MAXI all-sky survey Mid-band image 
(1.7$-$4.0 keV; Kimura et al. 2013).  Together with new evidence of a 
large amount of neutral matter absorbing the thermal plasma, in Paper-I \& II, we
argued that the observed  $kT$ $\simeq$ 0.3 keV gas was heated by a weak shock 
driven by the bubbles' expansion in the surrounding halo, 
with the corresponding velocity $v_{\rm exp}$ $\sim$ 300 km s$^{-1}$, 
which is consistent with the recent finding of a non-thermal velocity 
in the X-ray absorption line toward 3C~273 situated in the sightline 
of the Fermi bubbles (Fang \& Jiang 2014; but see also 
Fox et al. 2015 for the ultraviolet absorption line features toward PDS~456).
Such a low expansion velocity 
is also supported by some theoretical models discussing the Fermi 
bubbles' morphology (e.g., Crocker et al. 2014; Fujita et al. 2014;  
Mou et al. 2014). Also, Tahara et al. (2015) found possible 
evidence of 0.7 keV plasma in addition to 0.3 keV plasma 
in the northernmost region of the bubble.

While $kT$ $\simeq$ 0.3 keV plasma was ubiquitously observed in Papers-I and 
II, and was regarded as evidence of a shock-heated halo, these 
observations were highly biased toward the directions of X-ray enhancements and prominent 
structures like the NPS, N-cap and SE-claw. In fact, given the large spatial 
extent of the Fermi bubbles within the Galactic longitudes $|$$l$$|$ 
$<$ 20\deg\ and latitude $|$$b$$|$ $<$ 60\deg, most of the 
bubbles' interior were unprobed. 
Thus our goal in this paper is to 
determine the global characteristics and nature of diffuse X-ray emission 
associated with the Fermi bubbles, utilizing as many 
X-ray data pointings as possible.  We thus analyzed a total of 29 
archival datasets obtained with \textit{Suzaku} (Mitsuda et al.2007) 
and 68 archival datasets from \textit{Swift} (Gehrels et al. 2004) 
whose pointing centers are situated at Galactic longitudes $|$$l$$|$ $<$ 20\deg
and latitude 5\deg $\lesssim$ $|$$b$$|$ $<$ 60\deg, spanning the
full spatial extent of the Fermi bubbles above and below the GC. 
The observations and data reduction are described in section 2. 
The analysis process and results for \SZ and \SW are briefly 
summarized in section 3. 
In section 4, we discuss our findings in the context of proposed toy 
models assuming a (i) filled-halo without bubbles  
and a (ii) bubble-in-halo geometry. We also 
discuss a possible origin of asymmetry in the Galactic latitude profiles of the derived X-ray emission measure observed 
in the north and south 
bubbles. Section 5 presents our conclusions.

\begin{figure*}[t]
\begin{center}
\includegraphics[angle=0,scale=0.7]{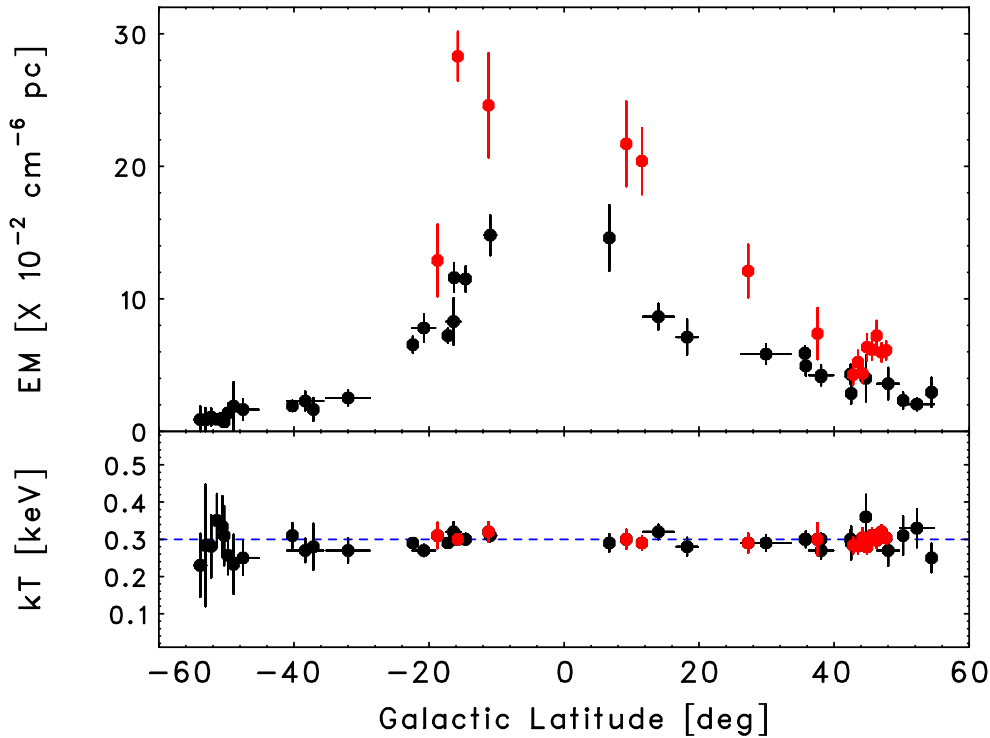}
\caption{Variation in the spectral fitting parameters EM ($top$) and $kT$ ($bottom$) for the APEC2 emission 
component as a function of Galactic latitude $b$. Abundances are fixed at $Z$ = 0.2 $Z_{\odot}$. 
The parameters determined for the NPS, SE-claw and NW-clump are shown 
in red (see Fig.~1). \label{fig:EMkT}}
\end{center}
\end{figure*}

\section{Observations and Data Reduction} 
\subsection{$Suzaku$ XIS}

As detailed in Paper-I and II, we conducted dedicated \SZ observations 
of the Fermi bubbles in 2012 and 2013 as a part of AO7 and and AO8 programs. 
The \textit{Suzaku} satellite (Mitsuda et al. 2007) is equipped with four 
X-ray telescopes (XRT; Serlemitsos et al. 2007) and each carries a 
focal-plane X-ray CCD camera (X-ray Imaging Spectrometer, XIS; 
Koyama et al. 2007a). One of the XIS sensors is a 
back-illuminated (BI) CCD (XIS1), and the other
three are front-illuminated (FI) ones (XIS0, XIS2, 
and XIS3). The field of view of \SZ XIS is 18'$\times$18' with a  
telescope half-power diameter (HPD, i.e., the point spread function) of 2'.
Since operation of XIS2 ceased in 2006 November due to 
contamination by a leakage current, we use only three CCDs in this 
paper. Although \textit{Suzaku}  also carries a hard X-ray detector 
(Takahashi et al. 2007),  we do not use the data collected by its PIN and GSO 
instruments because thermal emission we described below are too faint to 
be detected at above 10 keV and no statistically significant excess over the
cosmic X-ray background (CXB) were found with these PIN/GSO detectors. 
In the AO7 program (280 ksec total; Paper-I), eight pointings overlapped with the north-east bubble 
edge and across part of the NPS, with the remaining six pointings 
across the southernmost edges of the bubble. In AO8, we carried out four observations 
of 20 ksec each, pointed ``on'' and ``off'' the (i) N-cap and 
(ii) SE-claw regions (Paper-II). 

\begin{figure*}[t]
\begin{center}
\includegraphics[angle=0,scale=0.45]{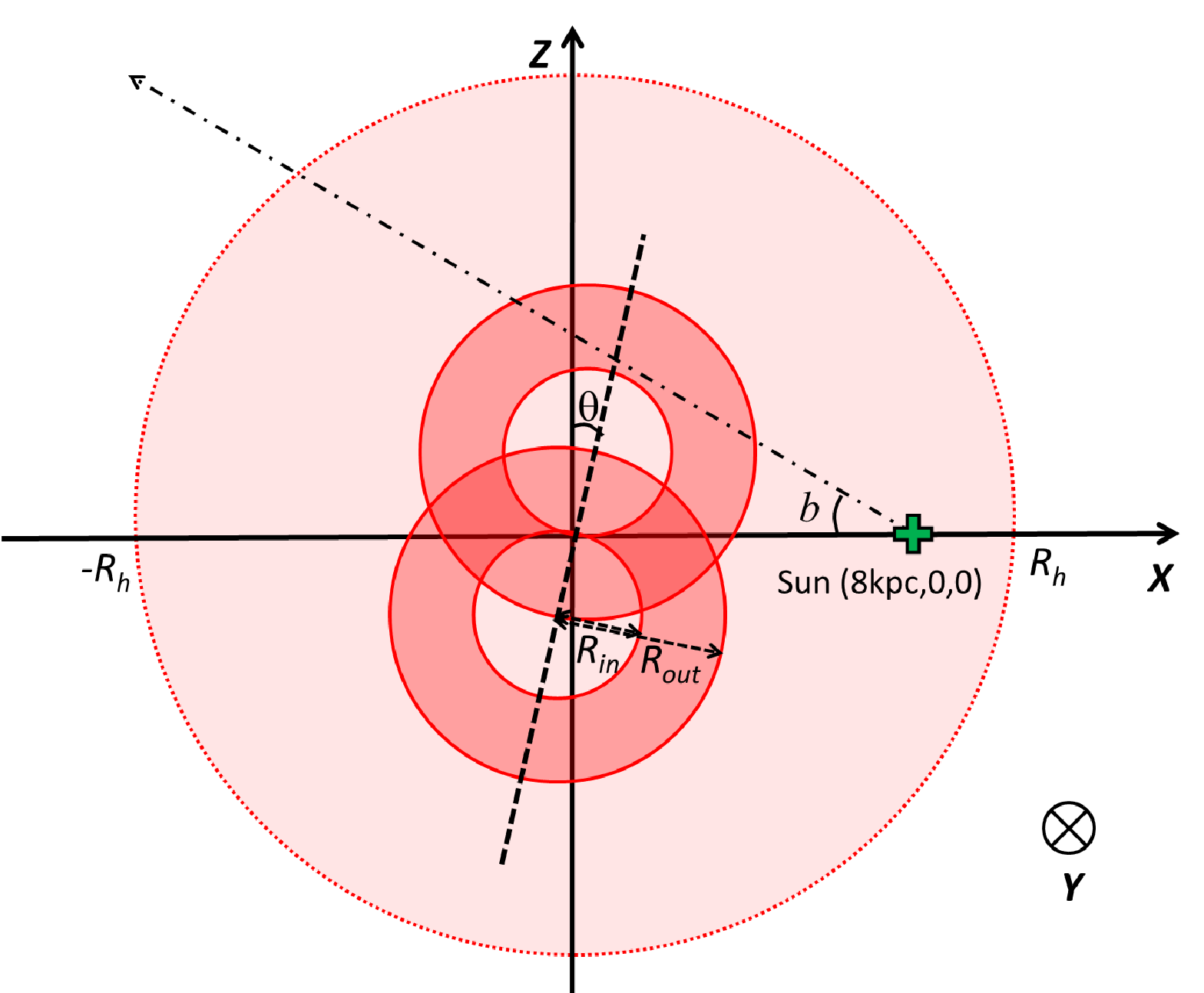}
\includegraphics[angle=0,scale=0.45]{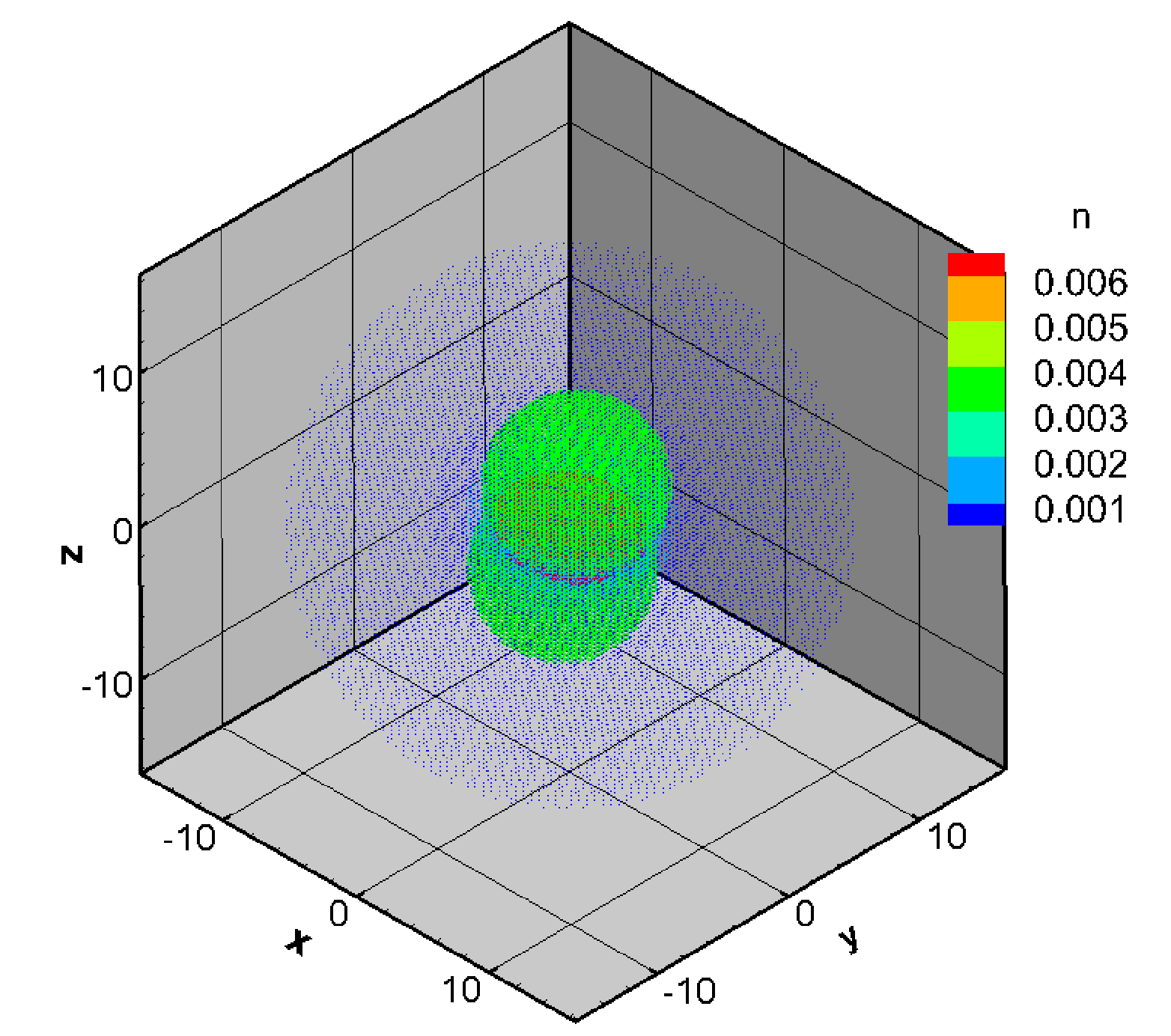}
\caption{A ``bubble-in-halo'' model assumed in this paper. As an underlying halo gas density profile, 
we assumed a $\beta$-model as detailed in the text. We set outer radius $R_{\rm out}$ = 5~kpc, 
inner radius $R_{\rm in}$ = 3~kpc, and inclination $\theta$ = 10$^{\circ}$.  $Top$: a cross sectional 
view at $l$ = 0$^{\circ}$. $Bottom$: A 3-D distribution of gas density profile $n(r)$ in units of cm$^{-3}$. \label{fig:geom}}
\end{center}
\end{figure*}

For this paper, we further investigated archived \textit{Suzaku} observations 
pointing toward the interior of the Fermi bubbles or in their close vicinity, 
covering $|$$l$$|$ $<$ 20$^{\circ}$ and 
$|$$b$$|$ $<$ 60$^{\circ}$. We selected pointings in which 
(i) the normal XIS observing mode was adopted throughout the observation, 
(ii) no bright X-ray features, such as compact sources and cluster gas, 
exist in the same filed of view that may affect the analysis of 
diffuse X-ray emission, and (iii) $|$$b$$|$ $\gtrsim$ 5\deg\ to 
avoid strong contamination from the GC region and/or bulge emission 
(e.g., Koyama et al. 2007b; Yuasa et al. 2012). 
A total of 29 \SZ pointings (14 from AO7, 4 from AO8 
and 11 from archival data) are analyzed in this paper. Note that five of these 
archival datasets are located near the N-cap area and
were published in Paper-II as ``N-cap1-5''. Table 1 summarizes 
all the times of the exposures and directions of the pointing centers of the 
\SZ datasets used in this paper. The \SZ pointing positions (focal centers) 
are overlaid as green or red circles onto the ROSAT 0.75 keV 
image in Fig.~1 with the boundary of the Fermi 
bubbles as drawn by Su et al. (2010) indicated.

We conducted all data reduction with the same methods as 
described in detail in Paper I \& II using the HEADAS software version 
6.14 and the calibration database (CALDB) released on 2013 August 13.
In summary, using \texttt{XSELECT}, the data corresponding to epochs of 
(i) low-Earth elevation angles (less than 20$^\circ$ 
during both night and day), (ii) the South Atlantic Anomaly 
(and 500 sec after) , and (iii) the low Cut-Off Rigidity 
(COR) of below 6 GV  were excluded.  
Hot and flickering pixels were removed 
using \texttt{SISCLEAN} (Day et al. 1998). Final images were created after the
Non X-ray Background (NXB) created with \texttt{XISNXBGEN} (Tawa et al. 2008) 
were subtracted from the raw XIS 0.4$-$10 keV images and a 
vignetting correction was applied using simulated flat sky images from \texttt{XISSIM} (Ishisaki et al. 2007). 

\subsection{$Swift$ XRT}

\SW (Gehrels et al. 2004) is an observatory mission whose 
primary goal is to explore and follow-up gamma-ray bursts. Its 
high mobility and sensitivity to localize sources 
especially using its X-ray Telescope (XRT; Burrows et al. 2005) 
makes it valuable for monitoring various X-ray sources within short 
exposures of typically $\le$ 5~ksec. 
The field of view of \SW XRT is 23.6'$\times$23.6' and the telescope 
HPD is 18'' at 1.5 keV.
While we did not conduct any dedicated 
\SW pointings of the Fermi bubbles as we did with \textit{Suzaku}, we found many 
short \SW pointings in the Fermi bubbles' direction, 
namely $|$$l$$|$ $<$ 20$^{\circ}$ and 
$|b|$ $<$ 60$^{\circ}$. Note that \SW also carries an ultraviolet/optical telescope (UVOT; Roming et al. 2005) and the
Burst Alert Telescope 
(BAT; Barthelmy et al. 2005), but we did not use 
these data because the thermal emission we describe below is too faint to be 
detected in the optical/ultraviolet and above 15 keV. 

We selected \SW observation pointings in which (i) no bright sources 
having XRT count rates of $\ge$ 0.6 cts s$^{-1}$ were found 
in the same field of 
view to avoid CCD pile up, and (ii) $|$$b$$|$ $>$ 5\deg\ to avoid 
contamination from the GC region and/or bulge emission.  
This selection yields 68 pointings which we analyzed in this paper. 
Note 19 of the \SW archival datasets located in the vicinity of the N-cap area were already analyzed in Paper-II 
as ``Swift1$-$19''. 
Table 2 summarizes the times of the exposures and directions of the pointing 
center of each \SW pointings used in this paper. The \SW pointing positions 
(focal centers) are indicated as green or red crosses in Fig.~1. Note, the six 
\SW pointings shown as red crosses exactly coincide with the X-ray 
enhancements / structures suggested to be associated with the 
Fermi bubbles, namely,  the NPS, SE-claw or NW-clump as shown in Fig.~1.

In the reduction of the \SW XRT data, the HEADAS software version 6.14 
and the  CALDB as of 2014 January 20 were used. 
In the XRT analysis, we only use the ``Photon Counting'' (PC)  mode data 
(Hill et al. 2004). We calibrated Level 1 data as recommended by 
{\it Swift} team\footnote{The \textit{Swift} XRT Data Reduction Guide:\\
\url{http://heasarc.nasa.gov/docs/swift/analysis/xrt\_swguide\_v1\_2.pdf}}. 
Specifically, we selected good time interval (GTI) from the Level 1 data using \texttt{xrtpipeline} and
the temperature of the CCDs were set to ``$\leqq -50$'' in the reduction.

\section{Analysis and Results}
\subsection{Extracting X-ray Spectra}

For the diffuse emission analysis of the \SZ data, we first ran the source 
detection algorithm in \texttt{XIMAGE} (Giommi et al. 1992) to eliminate
compact X-ray features from diffuse X-ray emission. We set the source region 
to the whole CCD chip that remained after excluding all the compact 
features detected at significance levels above 3\(\sigma\) with 
typical 2' radius circles enough to avoid the contamination from 
the compact sources. Then we used all the FI and BI CCDs, namely, XIS0, 1, 3 
for the spectral analysis to maximize the photon statistics. We made 
redistribution matrix files (RMFs) using \texttt{XISRMFGEN} (Ishisaki 
et al. 2007).  Auxillary response files (ARFs) were created 
using \texttt{XISSIMARFGEN} (Ishisaki 
et al. 2007) and new contamination files (released on 2013 August 13), 
assuming the uniform extension of the 
diffuse emission within 20' radii orbicular regions (giving the ARF area of 0.35 deg$^{2}$). We subtracted as background, the NXB data obtained from the region 
in the same CCD chip. Because some of the exposures are short
($\sim$10 ksec; Table~1), we carefully checked the analysis results 
by adopting different choices for source extraction radii and 
NXB/CXB models but the results were unchanged within the uncertainties 
given in Table 3.

\begin{figure*}[t]
\begin{center}
\includegraphics[angle=0,scale=0.8]{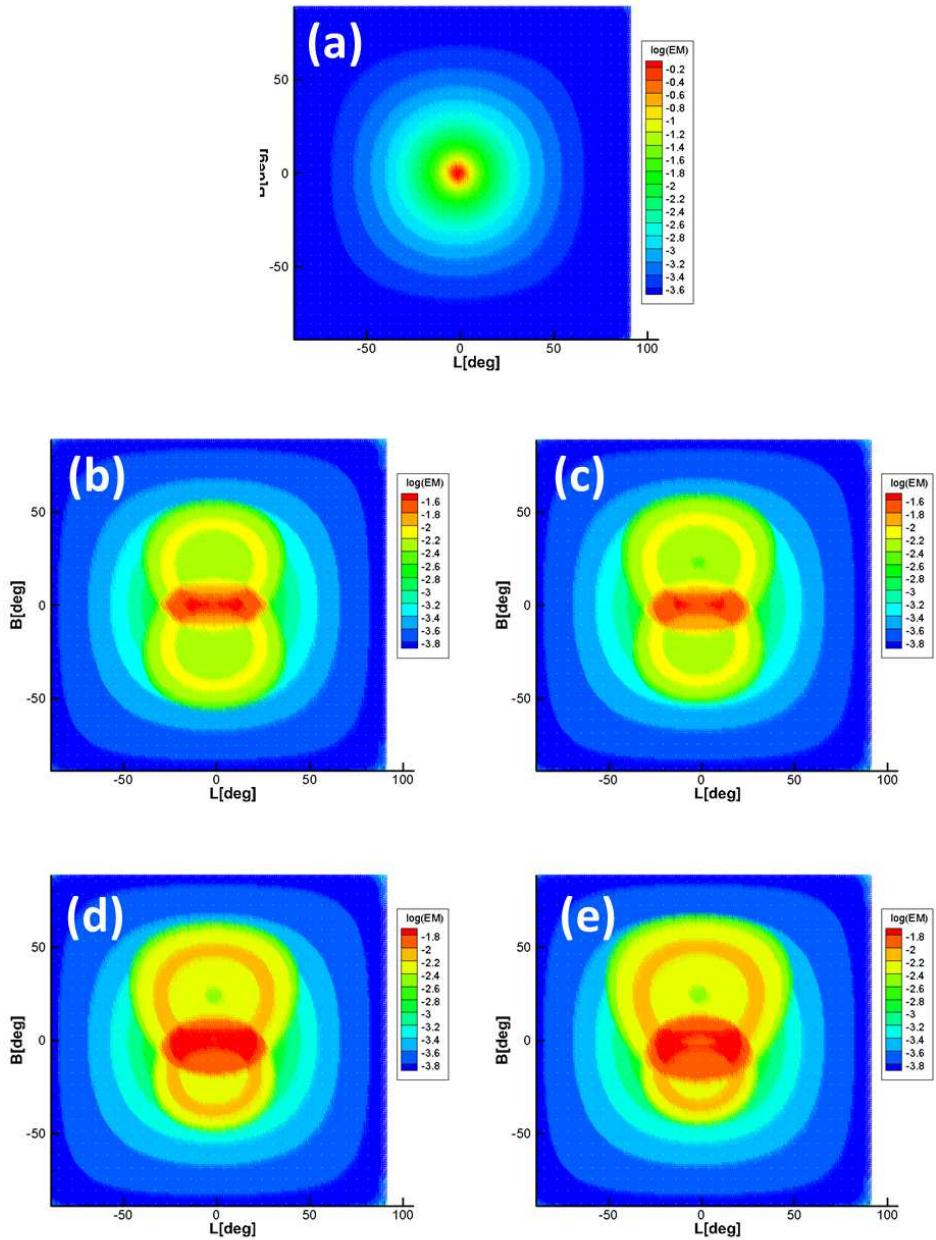}
\caption{Variation of EM in the ($l$, $b$) plane as observed from the Sun in the 
(a) filled halo model without bubbles, and bubble-in-halo models with (b) $\theta$ = 0\deg, (c) $\theta$ = 10\deg, (d) $\theta$ = 20\deg, and (e) $\theta$ = 30\deg.}
\end{center}
\end{figure*}

Similarly in the \SW XRT analysis, we extracted X-ray images in the 
energy range of 0.5$-$5~keV using \texttt{xselect}. Exposure maps were 
made using \texttt{xrtexpomap}. We ran the source detection algorithm 
in \texttt{XIMAGE}  and searched for X-ray compact features which 
were detected with photon statistics at $>$~3$\sigma$ confidence levels 
over the background.  In the XRT spectral analysis of the
diffuse emission, PHA files were extracted from event files 
with \texttt{xselect}. We made ARFs
using \texttt{xrtmkarf}, while we used the current redistribution matrix files 
(RMFs) in CALDB. 
To extract photons from diffuse X-ray emission only, we eliminated all the point sources using circles of 30\arcsec radius.

In contrast to \SZ data, evaluation of the instrumental background (NXB) is 
not well established for the \SW XRT data and studies are still ongoing 
(e.g., Moretti et al. 2009; 2011; 2012). However, as shown in
Moretti et al. (2011, Fig.~5 therein), the contribution of the NXB with respect to the CXB is less 
than 20$\%$ below 2~keV and gradually increases to $\gtrsim$50$\%$ at 
above 5~keV. Given that each \SW pointing (Table 2) is typically less 
than 10 ksec thus too short to derive meaningful spectra above 5 keV, 
we did not use the data above 5 keV for the spectral fitting. Moreover, 
we modeled the total XRT background as the sum of the NXB and CXB and checked 
that the analysis results for the diffuse emission were 
unchanged (within 1$\sigma$ uncertainty; see the next section) when changing the upper boundary 
to either 5~keV or 2~keV in the spectral fitting.

\subsection{Diffuse X-ray Emission}

Following Paper-I and II, all the spectra of the \SZ 
and \SW pointings after removing compact X-ray sources 
were fitted with a three component plasma model 
\textsc{apec1 + wabs*(apec2 + pl)} using \texttt{XSPEC}. 
The model consists of an \emph{unabsorbed} thermal component 
(denoted as \textsc{apec1}) which represent the Local Bubble 
emission and/or contamination from the Solar-Wind Charge 
Exchange (SWCX; Fujimoto et al. 2007), an \emph{absorbed} 
thermal component (denoted as \textsc{apec2}) representing the GH, 
and a single power-law component (denoted as \textsc{pl}) corresponding to the 
isotropic CXB radiation together with instrumental background for 
the case of \SW XRT. The photon index for the CXB component was fixed 
at $\Gamma_{\rm CXB} = 1.41$ (Kushino et al. 2002). The temperature and 
abundance of the LB plasma were fixed at $kT = 0.1$\,keV and $Z = Z_{\odot}$, 
respectively, as we did in Paper-I and II (see also, e.g., Yoshino et al. 
2009; Henley \& Shelton 2013). 

\begin{figure*}[t]
\begin{center}
\includegraphics[angle=0,scale=0.7]{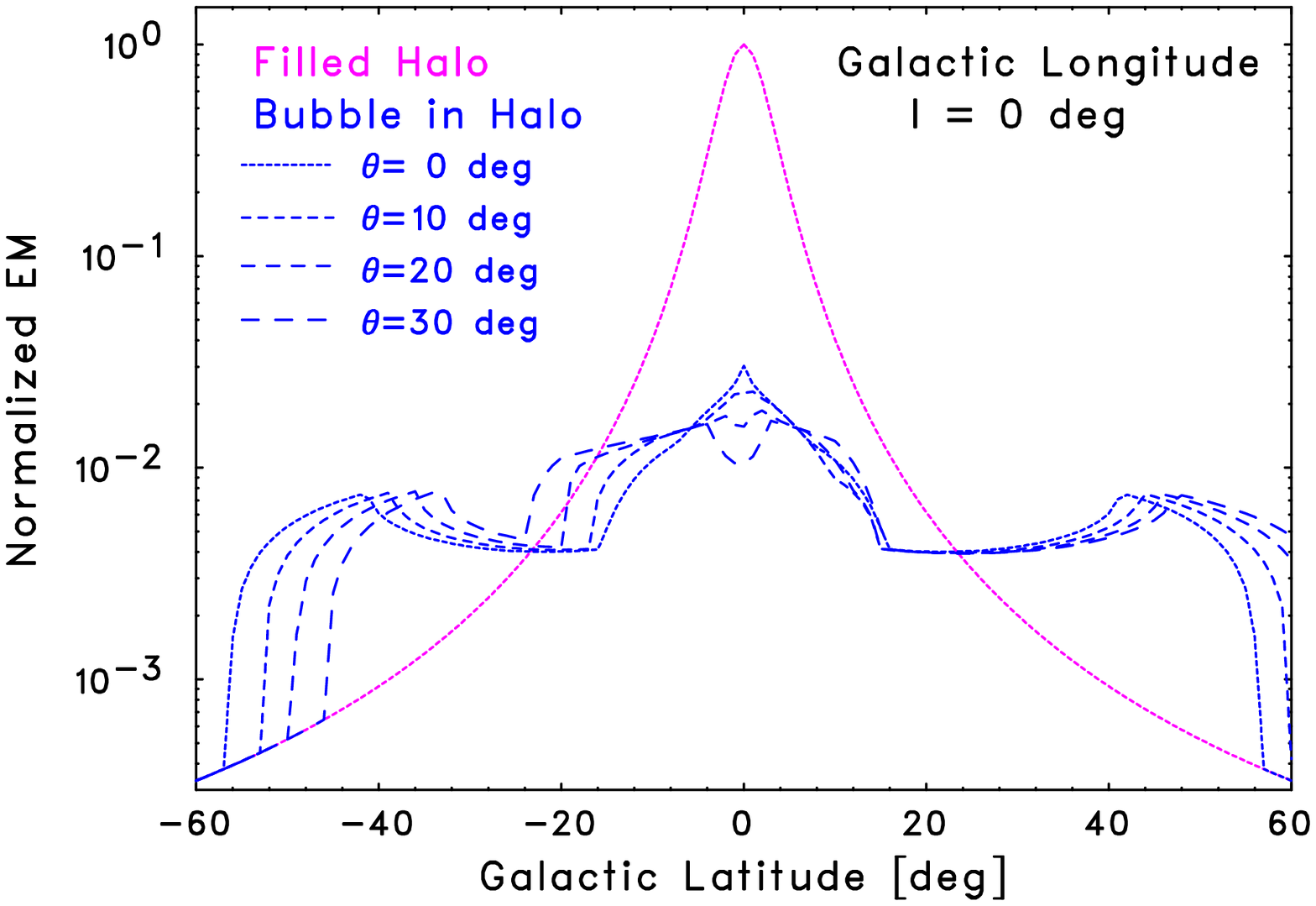}
\caption{Variation of EM as a function of Galactic latitude $b$ for (i) a filled halo model without bubbles and (ii) bubble-in-halo models as measured with $l$ = 0$^{\circ}$.
Different dashed lines correspond to inclination angles from $\theta$ = 0$^{\circ}$ to 30$^{\circ}$.}
\end{center}
\end{figure*}

As for the absorbed diffuse emission, the neutral hydrogen column density was 
fixed to the Galactic value $N_{\rm H, Gal}$ in the direction of each 
pointing because most of the values are consistent with the full Galactic 
values when $N_{\rm H}$ was left free in the spectral fitting. We also 
fixed the abundance of the \textsc{apec2} at $Z = 0.2 \, Z_{\odot}$, which 
is the on-average preferred value as detailed in Appendix B of Paper-I. 
Also this level of sub-solar metallicity is supported by a recent study 
of the GH using the $XMM$-$Newton$ Reflection Grating Spectrometer which 
measured the O VII K$_{\alpha}$ absorption line (Miller \& Bregman 2013; 
but see, e.g., Yao et al. (2005) and Yoshino et al. (2009) who assumed 
$Z = \, Z_{\odot}$). 
Even after reducing free parameters in the spectral fitting 
as described above, the photon statistics are too low to 
derive individual spectra for the 68 \SW XRT 
pointings, except 6 regions positioned at the bright X-ray enhancements denoted 
as the NPS, NW-clump, and SE-claw in Fig.~1. We therefore generated a spectrum 
by stacking \SW XRT data typically every 5\deg\ in Galactic latitude 
($\Delta$ $b$ $\simeq$ 5$-$15\deg) to increase the 
photon statistics. The results of our spectral fitting obtained for the \SZ 
data and \SW data are summarized in Table 3 and 4, respectively.  
In both tables, ``PL norm'' represents the power-law intensity as measured in 
2$-$10 keV, normalized by the absolute intensity of the CXB, namely 
$(5.85 \pm 0.38) \times 10^{-8}$\,erg\,cm$^{-2}$\,s$^{-1}$\,sr$^{-1}$ 
(Kushino et al. 2002). The value is close to unity for the \SZ data with 
some variations expected from the large-scale fluctuation of the CXB itself. 
The slightly larger values of PL norm in the \SW data indicate non-negligible contribution from 
the NXB in this energy band as mentioned above.

\subsection{EM and $kT$ distributions along Galactic latitude}

As can be seen in Tables 3 \& 4, the temperature of the GH as modeled by 
\textsc{apec2} is well represented by $kT$ $\simeq$ 0.3 keV, 
whilst the emission measure (EM) widely spans an order of 
magnitude depending on the Galactic latitude. To view the  
trend more clearly, Fig.~2 shows the variations of EM ($upper$) and $kT$ 
($bottom$) for the \textsc{apec2} emission component as a 
function of Galactic latitude $b$ for all the \SZ and \SW data. Red filled circles indicate X-ray 
enhancements corresponding to the
NPS, NW-clump and SE-claw, as also marked in red in Fig.~1. One can see the temperature is surprisingly 
uniform over a wide range of Galactic latitude  
5\deg $\lesssim$ $|$$b$$|$ $<$ 60\deg\ with fluctuations in $kT$ of 
only 0.30$\pm$0.07 keV over the whole spatial extent of the Fermi 
bubbles.

While the temperature values are uniform, the EM values increase steeply toward the GC (i.e., low $|$$b$$|$) 
with sudden jumps possibly related to the X-ray enhancements near 
the Fermi bubbles' edges. Moreover, the EM distribution is asymmetric with respect 
to the Galactic plane, 
decreasing more gradually in the north ($b$ $>$ 0\deg)
than in the 
south ($b$ $<$ 0\deg) toward high Galactic latitudes. 
For example, EM at 20\deg $<$ $b$ $<$ 35\deg, 
(5.82$\pm$0.76)$\times$10$^{-2}$ cm$^{-6}$ pc, is more than a factor of two 
larger than the corresponding EM in the south, where 
(2.52$^{+1.10}_{-0.52}$)$\times$10$^{-2}$ cm$^{-6}$ pc at  
$-$35\deg $<$ $b$ $<$ $-$25\deg. More about the origin of 
this asymmetry is discussed in the following section. 

\section{Discussion}

Following Paper-I \& II, we continued our systematic analysis of diffuse X-ray 
emission possibly related with the Fermi bubbles using data from both \SZ and \textit{Swift}. 
The X-ray data analyzed here were collected from archival observations 
covering Galactic longitude $|$$l$$|$ $<$ 20$^{\circ}$ and latitude 
5\deg $\lesssim$ $|$$b$$|$ $<$ 60\deg, approximately coinciding with 
the spatial extent of the Fermi bubbles. We showed that 
(i) the temperature of the GH is uniform along Galactic latitude with $kT$ $\simeq$ 
0.30$\pm$0.07 keV, (ii) the EM, in contrast, varies widely by more than an order 
of magnitude, with its values gradually decreasing toward high $b$, and 
(iii) the distribution of EM is asymmetric between the north and south bubbles.
While the north/south asymmetry is evident in the ROSAT 0.75 keV image 
(Snowden et al. 1995), we showed for the first time this is mainly accounted for by 
variations in the EM rather than differences in plasma temperature $kT$ that emits 
$\simeq$0.75 keV X-rays.
The observed $kT$ is a bit higher than what was derived for Galactic longitudes 
$65^{\circ}$ $<$ $l$ $<$ $295^{\circ}$ (Yoshino et al. 2009) and
$120^{\circ}$ $<$ $l$ $<$ $240^{\circ}$ (Henley et al. 2010; 
Henley \& Shelton 2013), regions that are well outside the bubbles' region, and hence was 
regarded as evidence of weak-shock heating during the bubble's expansion 
(Paper-I \& II). Although it is still unclear whether the observed $kT$ $\simeq$ 
0.3 keV plasma is really associated with the bubbles (see discussion in Paper-I), 
we are particularly interested in the global structure and 
asymmetry of EM ((ii) and (iii) described above) in order to further understand the  
possible relation between the observed $kT$ $\simeq$ 0.3 keV plasma with 
the Fermi bubbles. 

\begin{figure*}[t]
\begin{center}
\includegraphics[angle=0,scale=0.7]{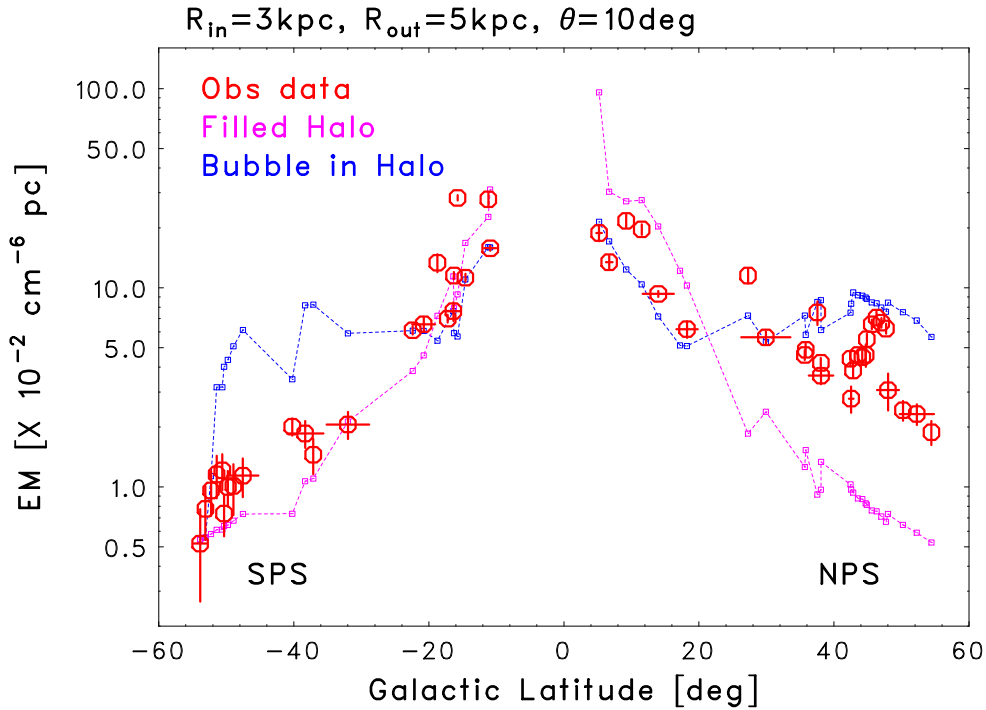}
\caption{Variations in the observed spectral fitting parameters EM for the APEC2 emission component (red) as a function of Galactic latitude $b$, compared with a toy model as shown in Figs.~3 \& 4. A larger fluctuation in the model line than in Fig.~4 is due to variations of $l$ for each observational pointings, ranging from
$-$20\deg $<$ $l$ $<$ 20\deg. Note that the profile from the ``bubble-in-halo'' model is consistent with the
north bubble data, while the filled halo model better represents the data for the south bubble (although note the
clear excess corresponding to the SPS).\label{fig:EMcomp}}
\end{center}
\end{figure*}

\subsection{A Model of the Bubbles in Galactic Halo}

Here we assume a simple model in which two spherical bubbles, 
that mimic the north and south Fermi bubbles, are embedded 
in the center of a gaseous halo with radius, $R_{\rm h}$ [kpc]. 
We set the GC at the origin of Cartesian space, and the Galactic disk is 
placed on the $xy$-plane with the Sun (i.e., observer) positioned  
at (8~kpc, 0, 0). 

As the underlying halo gas density profile, we assume a hydrostatic 
isothermal model (King profile or $\beta$ model; King 1962; 
Cavaliere \& Fusco-Femiano 1976) that follows
\begin{equation} 
n (r) = n_{\rm 0} (1 + (r/r_{\rm c})^2)^{-3 \beta/2}, 
\end{equation} 
where $n (r)$ is the gas density in cm$^{-3}$ at radius $r$ from the GC, 
$n_{\rm 0}$ is the density at $r$ = 0,  $r_{\rm c}$ is the core radius, and
$\beta$ is the slope of the profile at large radii.  Following recent studies of the 
structure of the GH based on X-ray data (e.g., Miller \& Bregman 2013), 
we hereafter set $r_{\rm c}$ = 0.5 kpc and $\beta$ = 2/3 
in this paper\footnote{More accurately,  Miller \& Bregman (2013) provided 
best-fit parameters $r_{\rm c}$ = 0.35$^{+0.29}_{-0.27}$ kpc and $\beta$ 
= 0.71$^{+0.13}_{-0.14}$. We thus set the values to rounded numbers within 
these uncertainties. Note,  $n(r)$ $\propto$ $r^{-2}$ for $r$ $\gg$ 
$r_{\rm c}$ when $\beta$ = 2/3.}. 
We also assume the halo boundary at $R_{\rm h}$ = 15 kpc for 
the purpose of this calculation.

We first calculated the EM profile of the GH $without$ 
bubbles for a direction of interest ($l$, $b$) from the Sun by
\begin{equation} 
EM(l,b) \propto \int n(r)^2 ds, 
\end{equation} 
where $ds$ is an element of length toward ($l$, $b$) direction 
(``filled-halo'' model). For comparison, we also considered a case in which two bubbles 
expand in the same halo by sweeping up surrounding halo gas 
(``bubble-in-halo'' model). We assume inner and outer radii of 
the bubbles, $R_{\rm in}$ and $R_{\rm out}$, 
where the centers of the northern and southern bubbles are 
positioned in the $xz$-plane (i.e., $y$ = 0). 
For simplicity, we assumed null gas density ($n$ = 0) 
inside each bubble, but the swept-up halo gas is distributed uniformly in 
shells with thickness $\Delta$$R$ = $R_{\rm out}$ $-$ $R_{\rm in}$, 
so that mass is conserved between two models. We remind the reader that a halo profile 
described above was first assumed by Miller \& Bregman (2013) based on the X-ray 
data $without$ considering bubbles, thus assuming the same profile both in the 
``filled-halo'' and ``bubble-in-halo'' models may be an oversimplification. 
Nevertheless, we show that our model can account for the global structure of 
isothermal diffuse X-ray emission as detailed below.
We also assumed an inclination of the northern and southern bubbles against the $z$-axis 
given by $\theta$.  Fig.~3 ($top$) shows a schematic 
view of the geometry assumed here (a cross-sectional view at 
$l$ = 0\deg\ and $\theta$ = 10\deg), and Fig.~3 ($bottom$) shows an example 
3-D diagram of the gas density profile $n(r)$ in our bubble-in-halo model.

Fig.~4 shows the variations of EM thus calculated in the ($l$, $b$) plane 
as observed from the Sun for a filled halo model without bubbles (a),   
and the bubble-in-halo model with various inclination angle from 
$\theta$ = 0\deg\ to 30\deg\ (b)-(e). Fig.~5 shows the corresponding variations of 
EM as a function of Galactic latitude $b$ 
in the case of a filled halo (magenta), and the bubble-in-halo models (blue), as measured for $l$ = 0\deg. 
We set $R_{\rm in}$ = 3~kpc and $R_{\rm out}$ = 5~kpc.
Note that EM is normalized to its peak value at $b$ = 0\deg\ of the filled 
halo model. 
In the absence of the bubbles, the filled halo model predicts a
sharp decrease of EM toward high Galactic latitudes, such that 
EM at $b$ = 60\deg\ is more than three orders of magnitude smaller 
than that derived at $b$ = 0\deg. In the case of the bubble-in-halo model, by contrast, there is more 
structure in the variations in EM which changes by only
about an order of magnitude.
Also one can see that the inclination $\theta$ may account for a 
certain degree of asymmetry in the EM, such that the northern bubble 
is spatially more extended toward high $b$ than the south bubble, as 
we see in Fig.4(e) for the case of $\theta$ = 30\deg. 
However, such a large inclination would similarly 
produce a high-degree of asymmetry in the gamma-ray bubbles, which 
strongly contradicts with the observations (e.g., Ackermann et al. 2014).

\subsection{Comparison with Data and Model: the North-South Asymmetry}

To determine to what extent the simple models described above can account 
for the observed EM profiles against $b$, we compared the model 
predictions to those determined from the observations. Since the observed $kT$ of the halo is 
uniform within the data analyzed here, we fixed $kT$ at 
0.30 keV and retried all the spectral fitting to reduce 
uncertainty in the EM values. Fig.~6 presents the thus obtained EM values (shown as red circles) 
compared with the predictions from the 
(i) filled-halo model without bubbles (magenta) and a  (ii) bubble-in-halo 
model assuming  $R_{\rm in}$ = 3~kpc, $R_{\rm out}$ = 5~kpc, and 
$\theta$ = 10\deg.  
Note that the vertical axis of Fig.~6 is shown on a logarithmic 
scale and the corresponding EM in the models were calculated 
from the same exact direction 
($l$, $b$) coincident with each observation
resulting in even larger fluctuations in the model line compared to that
shown in Fig.~5 (assuming $l$ = 0\deg) due to 
variations of $l$ for each observational pointing (that ranged from $-$20\deg $<$ $l$ $<$ 20\deg).
A gas density at the halo center corresponding to model 
lines shown in Fig.~6 is $n_0$ = 0.13 cm$^{-3}$ for the filled-halo 
model without bubbles, and the gas density 
in the shell is $n_{\rm shell}$ = 3.4$\times$10$^{-3}$ cm$^{-3}$ 
for the bubble-in-halo model which is doubled at low 
$b$ wherever the northern and southern shells overlapped 
(Fig.~3 $top$). Note that $n_{\rm shell}$ is almost consistent 
with what we observed for the NPS in Paper-I, namely, $n_g \simeq
4\times 10^{-3}$ cm$^{-3}$. Also, $n_0$ is consistent with 
that derived by Miller \& Bregman (2013), 
$n_0$ = 0.46$^{+0.74}_{-0.35}$ cm$^{-3}$, within the stated errors.

Even with the simple picture and geometry assumed here, 
our models qualitatively explain the observed EM profiles against 
$b$, although it appears that the observations in the north bubble
($b$ $>$ 0\deg) favor the (ii) bubble-in-halo model, whilst 
those of the south bubble ($b$ $<$ 0\deg) favor the (i) filled-halo 
model without bubbles. Observationally, this corresponds to the fact that 
such a bright and giant X-ray structure like the NPS is unseen in the 
south, which is often taken as evidence supporting the idea that the NPS 
and the rest of the Loop I structure arises from a nearby 
supernova remnant (see, detailed discussion in Paper-I).
However if we look at the 408 MHz radio map (Haslam et al. 1982; 
also Sofue 2000) 
closely, there is a southern counterpart of the NPS, ``South'' polar spur 
visible at $l$ $\sim$ 20\deg\ extending from ($l$, $b$) 
$\sim$ (20\deg, 0\deg) toward (30\deg, $-$30\deg), although it is
rather weak compared to the NPS (Sofue et al. 2000). Also, 
a western counterpart of the SPS, we call SPS-West, is found at 
($l$, $b$) $\sim$ (340\deg, 0\deg) to (320\deg, $-$30\deg).

Interestingly, in our X-ray data, we can also see a 
similar excess feature in the south against the filled-halo model 
at  $-$50\deg$<$ $b$ $<$ $-$30\deg, which is relatively symmetric 
with respect to the NPS, but this excess is small  
compared to the NPS (see, ``SPS'' in Figs.~1 \& 6). Here, the 
ratio of observed EM to the filled-halo model is $\gtrsim$~5 for the 
NPS whilst only $\lesssim$~2 in the SPS. As shown in Fig.~4, such 
a high degree of asymmetry in the north and south is difficult to explain 
solely by the inclination of bubbles' axis against the Galactic disk normal, 
thus suggesting an asymmetric outflow and/or initial density profile 
of the halo in which bubbles expand. 

The asymmetry of the NPS and SPS with respect to Galactic plane can 
be explained by both ``local'' and ``bubble'' models. Particularly as discussed 
in Paper-I, the NPS and the rest of the Loop I structure may be a nearby supernova remnants 
(SNR) located at a distance of 170 pc. Such an asymmetry, however, can also be 
explained by a large-scale outflow from the GC and  may not be exceptional in 
view of the fact that most 
shocked shells, such as supernova remnants and/or the GC phenomena, as well as 
extragalactic explosive events and bubbles, are more or less asymmetric 
like the NPS. An alternative model would be that the GH has a structural, 
as well as dynamical, asymmetry with respect to the Galactic plane and has an
axis caused by an intergalactic wind (Sofue 1994; 2000). If the Galaxy is 
moving towards the northeast, e.g., ($l$, $b$) $\sim$ (130\deg, 30\deg), where the 
warping of the HI gas disk is the highest observed, the northern halo will suffer from 
a stronger northeast wind of typically $\sim$100 km s$^{-1}$, whilst the 
southern halo is blocked from the wind by the Galactic disk. 
Such head/tail-winds to the bubbles and/or shocked shells could cause 
north-south (Galactic plane) as well as east-west (rotation axis) asymmetries 
in the sense that the north-east side is more enhanced, like in the NPS. 
Other more sophisticated modeling, including the  
intergalactic wind scenario, would be fruit subjects for future simulations.

In this context, one may also consider how the asymmetry of the NPS and SPS 
with respect to Galactic plane can be reconciled with the symmetric appearance of the gamma-ray 
bubbles observed with \F. If the former structures are physically associated 
with the bubbles, X-rays comes from swept-up gas of the surrounding halo $outside$ 
the bubbles that are clearly separated from the $inner$ bubbles that emit gamma-rays.
Thus according to various external/initial conditions of halo gas into which the bubbles 
expand, the X-ray envelope can be far from being symmetric as seen in gamma-rays.
Moreover, by analogy with extragalactic radio lobes (e.g., Scheuer 1995, and discussion therein), the bubble angles to the line of 
sight are not individually constrained by the symmetric appearance of the bubbles in gamma rays  (see also the case of the gamma-ray detection of the radio lobes of Cen~A; Abdo et al. 2010). The lines of sight adopted in the cartoon modeling span ranges adopted 
for extragalactic radio galaxies whose lobes also appear symmetric.

Although the global structures, metallicity, and density profile of 
the halo in our Galaxy is still under investigation 
(e.g., Miller \& Bregman 2013), future extensive 
studies using the MAXI-SSC (Matsuoka et al. 2009; Tsunemi et al. 2010) 
and $Astro$-$H$ (Takahashi et al. 2014) will further clarify the 
origin, interaction, and dynamics between the hot gas halo and the 
bubbles. Particularly $Astro$-$H$, the sixth X-ray astronomy mission in 
Japan, carries the Soft X-ray Spectrometer 
(SXS; Mitsuda et al. 2014) which provides the capability for high resolution 
X-ray spectroscopy with $<$ 7eV (FWHM) in the energy range of 0.3$-$10~keV. 
In this context, Fox et al. (2015) reported two high-velocity metal 
absorption components centered at $v_{\rm LSR}$ = $-$235 and +250 
km s$^{-1}$ from ultraviolet spectra, 
which can be explained with an outflow velocity of $\gtrsim$ 900 km s$^{-1}$ and a full 
opening angle of $\simeq$110\deg. While the velocity is higher than 
in Paper-I \& II, such a value depends on the geometry of the biconical outflow 
assumed in the model. In this context, we note again that a slower velocity 
$v_{\rm exp}$ $\sim$ 300 km s$^{-1}$, which is consistent with Paper-I \& II is 
implied by the X-ray absorption line toward 3C~273. Moreover, the presence of another 
$kT$ $\simeq$ 0.7 keV plasma, corresponding to $v_{\rm exp}$ $\sim$ 600 km s$^{-1}$,  
is reported in Paper-II.
As discussed in detail in Inoue et al. (2015), precise 
measurements of metal abundances in the halo gas will provide crucial hints 
for the origin of the Fermi bubbles, either from the past activity of a GC-like 
AGN or nuclear starforming activity. As $Astro$-$H$ will be launched 
in the winter of 2015, this will enable further progress toward 
clarifying the Fermi bubbles' nature.

\section{Conclusion}

In this paper we presented a systematic analysis of X-ray data provided by  
\textit{Suzaku} (29 pointings) and \textit{Swift} (68 pointings), 
covering sightlines through the entire spatial extent of the Fermi bubbles. 
We showed that (i) the temperature of the GH is surprisingly uniform 
with Galactic latitude with $kT$ $\simeq$ 0.30$\pm$0.07 keV, 
(ii) the EM, in contrast, varies widely by more than an order 
of magnitude, gradually decreasing toward high $b$, and 
(iii) that the distribution of EM is asymmetric between the north and south bubbles.
Although the association of the X-ray emission with the bubbles is not conclusive, 
we compared our observations with simple models 
assuming (i) a filled halo without bubbles, whose gas density follows 
a hydrostatic isothermal $\beta$ model and (ii) 
a bubble-in-halo in which two identical bubbles expand 
within a halo forming a thick uniform shell of swept-up halo gas. 
We showed that a weak X-ray excess feature against filled-halo model, the 
SPS, is evident in the south, but is rather weak compared to the NPS.  
Such a high degree of asymmetry is difficult to  
explain only by the effect of an inclined axis of the bubbles. This may suggest
an asymmetric outflow and/or anisotropic initial density profile in-situ, although 
this is inconclusive based on the current X-ray data presented in this paper.

\acknowledgments 

We acknowledge the referee for useful suggestions that improved the 
manuscript. Work by C.C.C. at NRL is supported in part by NASA DPR S-15633-Y.

\begin{deluxetable*}{lllrrrrrl}
\tabletypesize{\scriptsize}
\tablecaption{\SZ observation log}
\tablewidth{0pt}
\tablehead{
\colhead{ID} & \colhead{Start time} & \colhead{Stop time} & \colhead{R.A.} & \colhead{Decl.} & \colhead{$l$} & \colhead{$b$} &  \colhead{Exposure} & \colhead{Note$^f$}\\
\colhead{} & \colhead{(UT)} & \colhead{(UT)} & \colhead{[$^{\circ}$]$^a$} & \colhead{[$^{\circ}$]$^b$} & \colhead{[$^{\circ}$]$^c$} & \colhead{[$^{\circ}$]$^d$} &  \colhead{[ksec]$^e$} & \colhead{}
}
\startdata
\multicolumn{9}{c}{North bubble} \\
\tableline
    507006010 & 2012/08/08 10:23 & 2012/08/08 23:03 & 233.401 &  9.076  & 15.480 & 47.714 & 17.5 (45.5) & I (N1)\\  
    507005010 & 2012/08/07 23:41 & 2012/08/08 10:22 & 233.623 &  8.079  & 14.388 & 47.011 & 16.2 (36.9) & I (N2)\\  
    507004010 & 2012/08/07 10:31 & 2012/08/07 23:40 & 233.834 &  7.087  & 13.321 & 46.308 & 17.6 (46.2) & I (N3)\\  
    507003010 & 2012/08/06 23:20 & 2012/08/07 10:30 & 234.034 &  6.098  & 12.280 & 45.606 & 16.7 (40.2) & I (N4)\\  
    507001010 & 2012/08/05 23:04 & 2012/08/06 09:33 & 234.250 &  5.090  & 11.255 & 44.871 & 15.3 (36.0) & I (N5)\\  
    507002010 & 2012/08/06 09:34 & 2012/08/06 23:18 & 234.405 &  4.131  & 10.263 & 44.204 & 19.0 (47.9) & I (N6)\\  
    507007010 & 2012/08/08 23:06 & 2012/08/09 10:20 & 234.551 &  3.174  &  9.291 & 43.537 & 17.0 (40.4) & I (N7)\\  
    507008010 & 2012/08/09 10:21 & 2012/08/09 23:53 & 234.713 &  2.200  &  8.334 & 42.838 & 12.0 (24.9) & I (N8)\\
    508007010 & 2013/07/26 08:09 & 2013/07/26 20:11 & 221.750 &  -1.312 & 351.952 &  50.223 &  20.7 (40.7) & II (N\_cap\_on)\\
    508008010 & 2013/07/26 20:16 & 2013/07/27 10:14 & 233.686 &  -9.893 & 355.509 &  35.809 & 19.6 (48.8) & II (N\_cap\_off)\\
    807062010 & 2012/08/01 23:39 & 2012/08/02 10:54 & 217.761 &   0.794 & 349.311 &  54.438 & 15.3 (40.4) & II (N\_cap\_1)\\
    807058010 & 2012/07/28 08:10 & 2012/07/28 17:58 & 233.434 &   3.616 &   8.894 &  44.702 & 10.4 (38.8) & II (N\_cap\_2)\\
    705026010 & 2011/02/01 18:51 & 2011/02/02 04:25 & 230.255 &  -3.837 & 358.141 &  42.451 & 17.5 (31.7) & II (N\_cap\_3)\\
    701079010 & 2006/07/19 17:39 & 2006/07/20 15:02 & 220.569 & -17.330 & 337.266 &  38.061 & 32.0 (71.1) & II (N\_cap\_4)\\
    401001040 & 2006/02/27 20:38 & 2006/02/28 23:00 & 226.648 & -16.180 & 344.020 &  35.677 & 28.7 (94.4) & II (N\_cap\_5)\\
\tableline
\multicolumn{9}{c}{South bubble} \\
\tableline
    507013010 & 2012/04/19 14:11 & 2012/04/20 02:44 & 332.668 & -46.192 & 351.010 & -53.100 &  18.1 (41.2) & I (S1)\\
    507012010 & 2012/04/19 03:15 & 2012/04/19 14:10 & 331.474 & -46.348 & 351.149 & -52.265 &  11.5 (38.8) & I (S2)\\        
    507010010 & 2012/04/18 04:59 & 2012/04/18 16:10 & 330.278 & -46.492 & 351.281 & -51.432 &  11.2 (38.8) & I (S3)\\        
    507009010 & 2012/04/17 16:40 & 2012/04/18 04:58 & 329.080 & -46.624 & 351.406 & -50.602 &  21.0 (42.5) & I (S4)\\        
    507011010 & 2012/04/18 16:12 & 2012/04/19 03:12 & 327.882 & -46.743 & 351.525 & -49.775 &  18.1 (36.9) & I (S5)\\        
    507014010 & 2012/04/20 02:47 & 2012/04/20 14:25 & 326.683 & -46.851 & 351.638 & -48.950 &  11.1 (40.2) & I (S6)\\        
    508009010 & 2013/04/22 16:51 & 2013/04/23 07:56 & 287.398 & -27.250 &   9.973 & -15.747 &  11.8 (48.8) & II (SE\_on)\\
    508010010 & 2013/04/23 07:58 & 2013/04/23 19:59 & 288.748 & -25.775 &  11.875 & -16.290 &  16.0 (43.1) & II (SE\_off)\\
    500003010 & 2006/03/08 17:41 & 2006/03/09 01:07 & 282.688 & -33.893 &   1.999 & -14.596 &  9.89 (25.2) & BULGE\_6\\
    100041020 & 2006/03/23 22:31 & 2006/03/25 10:38 & 284.147 & -37.910 &  -1.403 & -17.211 &  63.5 (129) & RXJ1856\\
    705014010 & 2010/04/13 06:37 & 2010/04/14 00:08 & 285.522 & -51.170 & -14.421 & -22.401 &  23.0 (58.7) & EMS1274\\
    705028010 & 2010/10/28 10:57 & 2010/10/29 03:19 & 309.873 & -56.354 & -18.821 & -37.128 &  15.9 (58.7) & EMS1388\\ 
    806079010 & 2011/05/08 23:24 & 2011/05/10 00:21 & 319.721 & -63.575 & -29.263 & -40.234 &  32.9 (83.2) & RCS2118\\
    703012010 & 2008/05/11 12:28 & 2008/05/12 13:35 & 327.081 & -34.951 &  10.029 & -50.337 &  32.7 (89.1) & NGC7130\\
\tableline
\enddata 
\tablecomments{$^a$: Right ascension of \SZ pointing center in J2000 equinox.\\
$^b$: Declination of \SZ pointing center in J2000 equinox.\\
$^c$: Galactic longitude of \SZ pointing center.\\
$^d$: Galactic latitude of \SZ pointing center.\\
$^e$: \SZ XIS exposure in ksec  that was actually used in the analysis, as compared with total elapsed time for the observation shown in parenthesis.\\
$^f$: Reference or focusing target of the observations. I and II denote the data from Paper-I and Paper-II, respectively, that were uniformly 
reanlyzed here. N1$-$8, S1$-$6, N\_cap\_on, 
N\_cap\_on, SE\_on and SE\_off denote dedicated observations for the Fermi bubbles conducted in \SZ AO7 and AO8 from Papers-I and -II, while the remaining 
are newly analyzed archival datasets.
}
\end{deluxetable*}

\begin{deluxetable*}{lllrrrrrr}
\tabletypesize{\scriptsize}
\tablecaption{\SW observation log}
\tablewidth{0pt}
\tablehead{
\colhead{ID} & \colhead{Start time} & \colhead{Stop time} & \colhead{R.A.} & \colhead{Decl.} & \colhead{$l$} & \colhead{$b$} &  \colhead{Exposure} & \colhead{Note$^f$}\\
\colhead{} & \colhead{(UT)} & \colhead{(UT)} & \colhead{[$^{\circ}$]$^a$} & \colhead{[$^{\circ}$]$^b$} & \colhead{[$^{\circ}$]$^c$} & \colhead{[$^{\circ}$]$^d$} &  \colhead{[ksec]$^e$} & \colhead{}
}
\startdata
\multicolumn{9}{c}{North bubble} \\

\tableline
      00037600001 & 2009/12/19 01:45 & 2009/12/19 08:34 & 217.296 &   1.301 & 349.261 & 55.125 &  5.42 (24.7) & II (Swift1)\\
      00037755001 & 2009/12/18 03:11 & 2009/12/18 16:14 & 216.922 &   0.005 & 347.337 & 54.359 &  7.59 (47.0) & II (Swift2)\\
      00091308008 & 2013/03/26 06:42 & 2013/03/26 11:56 & 217.608 &  -1.829 & 346.343 & 52.493 &  4.57 (18.9) & II (Swift3)\\
      00082093002 & 2013/09/12 14:18 & 2013/09/12 22:12 & 224.263 &   2.802 & 359.375 & 51.352 &  2.62 (28.5) & II (Swift4)\\
      00032864005 & 2013/06/30 05:36 & 2013/06/30 18:46 & 224.975 &   1.894 & 359.016 & 50.216 &  5.49 (47.4) & II (Swift5)\\
      00033207001 & 2014/03/28 00:14 & 2014/03/28 17:27 & 225.232 &   1.908 & 359.351 & 50.035 &  8.72 (62.0) & II (Swift6)\\
      00033265002 & 2014/04/28 10:56 & 2014/04/28 12:39 & 226.089 &   2.310 &   0.777 & 49.575 &  2.98 (6.17) & II (Swift7)\\
      00090306002 & 2010/12/25 16:19 & 2010/12/25 18:07 & 227.614 &   1.755 &   1.577 & 48.145 &  1.56 (6.51) & II (Swift8)\\
      00090281001 & 2010/04/08 10:32 & 2010/04/08 13:55 & 226.089 &  -2.597 & 355.346 & 46.306 &  2.17 (12.2) & II (Swift9)\\
      00036338003 & 2008/01/07 13:54 & 2008/01/07 23:44 & 227.732 &  -5.725 & 353.912 & 42.925 &  5.01 (35.4) & II (Swift10)\\
      00039721001 & 2010/12/29 00:45 & 2010/12/30 17:04 & 234.249 &   0.959 &   6.602 & 42.472 &  5.14 (145) & II (Swift11)\\
      00037942001 & 2008/06/22 07:20 & 2008/06/22 14:02 & 233.223 &  -0.749 &   3.927 & 42.230 &  5.21 (24.2) & II (Swift12)\\
      00039723001 & 2011/01/02 01:06 & 2011/01/02 06:16 & 235.090 &  -2.041 &   4.141 & 39.966 &  5.26 (18.6) & II (Swift13)\\
      00040980001 & 2010/09/16 01:01 & 2010/09/16 07:44 & 228.007 & -10.863 & 349.572 & 38.957 &  5.44 (24.2) & II (Swift14)\\
      00055750014 & 2011/01/14 03:35 & 2011/01/14 07:05 & 232.086 &  -7.240 & 356.528 & 38.790 &  2.56 (12.7) & II (Swift15)\\
      00035800002 & 2006/10/08 16:10 & 2006/10/09 05:09 & 245.412 &   9.558 &  23.816 & 37.542 &  4.58 (46.8) & II (Swift16)\\
      00037281001 & 2008/01/20 00:44 & 2008/01/20 16:59 & 241.792 &   1.112 &  12.405 & 36.399 &  8.69 (58.5) & II (Swift17)\\
      00037279002 & 2008/12/26 00:03 & 2008/12/27 22:53 & 224.853 & -16.693 & 341.960 & 36.300 &  8.52 (169) & MASER1459\\
      00036065002 & 2007/01/17 00:55 & 2007/01/17 23:33 & 236.199 & -11.491 & 356.179 & 32.920 &  6.16 (81.5) & II (Swift18)\\
      00038072003 & 2010/01/07 10:10 & 2010/01/07 22:58 & 235.510 & -14.168 & 353.354 & 31.538 &  7.87 (46.1) & J1542\\
      00041776003 & 2011/03/27 08:19 & 2011/03/27 19:52 & 224.702 & -24.950 & 336.289 & 29.547 &  7.05 (41.6) & J1458\\
      00037283001 & 2008/01/20 18:24 & 2008/01/21 13:46 & 253.272 &   2.403 &  20.746 & 27.269 &  8.03 (69.7) & II (Swift19)\\
      00037188002 & 2008/01/16 00:15 & 2008/01/16 22:59 & 247.263 &  -9.882 &   5.589 & 25.601 &  10.7 (81.8) & J1629\\
      00090500002 & 2010/07/06 04:13 & 2010/07/06 12:36 & 243.873 & -22.205 & 353.022 & 20.248 &  5.72 (30.2) & UKSCE-1\\ 
      00046310001 & 2013/01/31 01:37 & 2013/01/31 08:07 & 252.202 & -17.317 &   2.269 & 17.310 &  4.30 (23.4) & PBCJ1648\\
      00036649002 & 2007/10/08 03:07 & 2007/10/08 14:31 & 249.626 & -20.944 & 357.709 & 17.010 &  4.59 (41.1) & IGRJ1638\\
      00041223001 & 2010/09/28 08:24 & 2010/09/28 15:27 & 250.605 & -22.371 & 357.144 & 15.407 &  4.72 (25.4) & IGRJ1642\\
      00035086002 & 2007/02/24 00:06 & 2007/02/24 14:40 & 262.590 & -5.9926 &  17.929 & 15.013 &  12.7 (52.8) & IGRJ1730\\
      00037644001 & 2009/02/24 10:48 & 2009/02/24 17:16 & 250.075 & -23.896 & 355.599 & 14.827 &  3.46 (23.3) & HD150193\\
      00090182002 & 2010/01/23 11:43 & 2010/01/23 23:06 & 253.660 & -19.269 &   1.496 & 15.034 &  3.98 (41.0) & J1654\\
      00038075002 & 2010/01/23 02:04 & 2010/01/23 10:15 & 246.613 & -29.856 & 348.871 & 13.260 &  4.67 (29.5) & J1626\\
      00090991002 & 2011/02/02 04:01 & 2011/02/02 23:34 & 252.873 & -26.009 & 355.535 & 11.526 &  8.90 (70.4) & AS210\\
      00036347001 & 2007/02/27 00:16 & 2007/02/27 15:02 & 263.261 & -13.080 &  12.032 & 10.812 &  10.7 (53.0) & MOJ2B1730\\
      00035348002 & 2006/02/03 00:04 & 2006/02/03 22:53 & 252.047 & -30.599 & 351.430 &  9.223 &  9.21 (82.1) & IGRJ1648\\
      00036118001 & 2007/01/27 16:12 & 2007/01/28 00:23 & 252.505 & -33.116 & 349.710 &  7.330 &  4.60 (29.5) & IGRJ1650\\
      00035647002 & 2007/02/06 01:23 & 2007/02/06 23:59 & 253.794 & -33.162 & 350.355 &  6.460 &  6.95 (81.4) & J1655\\
      00035272002 & 2006/06/13 16:39 & 2006/06/13 21:50 & 254.072 & -33.079 & 350.567 &  6.330 &  4.79 (18.7) & J1656\\
      00037646002 & 2010/11/02 03:49 & 2010/11/02 05:42 & 266.309 & -17.946 &   9.364 &  5.779 &  1.88 (6.80) & GLMP632 \\
      00036121001 & 2007/02/27 16:20 & 2007/02/27 23:03 & 263.283 & -24.113 &   2.606 &  4.928 &  6.12 (24.2) & IGRJ1733\\
      00031277001 & 2008/10/16 06:10 & 2008/10/16 23:55 & 265.538 & -20.916 &   6.435 &  4.861 &  4.35 (63.9) & J1741\\
\tableline
\multicolumn{9}{c}{South bubble} \\
\tableline
      00091760004 & 2013/11/06 02:50 & 2013/11/06 11:02 & 272.290 & -41.224 & 351.638 & -10.236 &  3.79 (29.5) & AS276\\
      00031677002 & 2010/11/03 08:36 & 2010/11/03 23:27 & 282.418 & -23.811 &  11.316 & -10.242 &  3.79 (53.5) & ROSS154\\
      00090992004 & 2010/11/06 04:06 & 2010/11/06 20:17 & 283.279 & -24.328 &  11.178 & -11.174 &  5.07 (58.3) & AS327\\
      00048048002 & 2012/05/06 03:53 & 2012/05/07 17:05 & 282.008 & -26.841 &   8.363 & -11.191 &  3.36 (134) & PBCJ1847\\
      00036632002 & 2007/08/05 08:26 & 2007/08/05 19:48 & 281.304 & -30.254 &   4.933 & -12.056 &  5.12 (41.0) & J1845\\
      00035794001 & 2007/06/19 17:48 & 2007/06/19 22:52 & 276.781 & -46.941 & 347.751 & -15.594 &  3.37 (18.3) & XMMSL1J1827\\
      00036405001 & 2008/05/30 08:56 & 2008/05/31 23:45 & 288.888 & -24.179 &  13.456 & -15.786 &  7.15 (140) & HD1799\\
      00036289001 & 2007/04/08 01:13 & 2007/04/08 09:25 & 274.940 & -55.356 & 339.182 & -17.784 &  3.15 (29.6) & J1819\\
      00040716003 & 2010/08/25 00:45 & 2010/08/25 12:13 & 289.868 & -29.974 &   8.178 & -18.777 &  4.47 (41.3) & PBCJ1919\\
      00035839001 & 2007/04/27 11:09 & 2007/04/27 19:32 & 284.035 & -43.056 & 353.500 & -18.944 &  4.06 (30.2) & XMMSL1J1856\\
      00038080002 & 2008/11/02 01:10 & 2008/11/02 11:12 & 279.767 & -57.281 & 338.240 & -20.958 &  8.35 (36.1) & SWIFTJ1839\\
      00031727001 & 2010/05/26 10:13 & 2010/05/26 15:13 & 285.522 & -51.170 & 345.578 & -22.404 &  4.24 (18.0) & 1FGLJ1902\\
      00041100002 & 2010/06/11 05:10 & 2010/06/11 23:09 & 294.536 & -51.136 & 346.988 & -27.909 &  7.03 (64.8) & SWIFTJ1938\\
      00032516006 & 2012/07/22 14:55 & 2012/07/23 06:57 & 305.912 & -28.278 &  14.862 & -31.529 &  3.92 (57.8) & PSNJ2023\\
      00037330002 & 2008/06/18 01:24 & 2008/06/18 23:59 & 304.610 & -55.650 & 342.270 & -34.232 &  5.82 (81.3) & SWIFTJ2018\\
      00041108001 & 2010/12/02 06:32 & 2010/12/02 21:16 & 308.602 & -30.602 &  12.905 & -34.391 &  7.39 (53.0) & SWIFTJ2034\\
      00035790004 & 2007/03/30 00:05 & 2007/03/30 08:41 & 307.684 & -48.788 & 350.669 & -36.101 &  3.78 (31.0) & XMMSL1J2030\\
      00041479002 & 2011/02/21 02:29 & 2011/02/21 11:59 & 309.873 & -56.354 & 341.182 & -37.125 &  3.61 (34.2) & 1FGLJ2039\\
      00046327002 & 2012/06/20 01:54 & 2012/06/20 23:01 & 310.648 & -53.695 & 344.465 & -37.817 &  3.85 (76.0) & PBCJ2042\\
      00080269001 & 2013/07/08 07:35 & 2013/07/08 16:12 & 313.008 & -57.069 & 339.991 & -38.735 &  7.12 (31.1) & PBCJ2052\\
      00091684001 & 2013/04/02 01:19 & 2013/04/02 23:51 & 313.072 & -57.064 & 339.991 & -38.770 &  4.78 (81.2) & SWIFTJ2052\\
      00041188004 & 2011/07/02 23:29 & 2011/07/03 22:21 & 319.007 & -58.662 & 337.033 & -41.490 &  5.12 (82.4) & SWIFTJ2116\\
      00035232001 & 2005/12/07 00:22 & 2005/12/07 23:10 & 320.308 & -43.007 & 358.079 & -44.971 &  9.68 (82.1) & SWIFTJ2121\\
      00033015009 & 2014/04/01 14:34 & 2014/04/01 23:01 & 324.363 & -47.032 & 351.833 & -47.361 &  4.29 (30.4) & ESO287\\
      00038411002 & 2009/04/05 14:50 & 2009/04/06 07:01 & 324.850 & -42.589 & 358.318 & -48.326 &  6.41 (58.3) & MH2136\\
      00039206001 & 2009/09/22 08:10 & 2009/09/22 13:26 & 326.255 & -33.955 &  11.447 & -49.629 &  6.58 (19.0) & PMNJ2145\\
      00037292001 & 2008/04/06 06:55 & 2008/04/07 15:12 & 330.321 & -37.773 &   5.315 & -52.906 &  11.1 (116) & MASER2201\\
      00040395004 & 2012/09/25 04:02 & 2012/09/25 23:31 & 335.239 & -46.036 & 350.319 & -54.843 &  9.51 (70.2) & IC5201\\
\enddata 
\tablecomments{$^a$: Right ascension of \SW pointing center in J2000 equinox.\\
$^b$: Declination of \SW pointing center in J2000 equinox.\\
$^c$: Galactic longitude of \SW pointing center.\\
$^d$: Galactic latitude of \SW pointing center.\\
$^e$: \SW XRT exposure in ksec that was actually used in the analysis, as compared with total elapsed time for the observation shown in parenthesis.\\
$^f$: Reference or focusing target. II denote data presented in Paper- II and uniformly reanalyzed here while the rest
are newly presented in this paper. 
}
\end{deluxetable*}


\begin{deluxetable*}{lccccccc}
\tabletypesize{\scriptsize}
\tablecaption{Fitting parameters for \SZ observations}
\tablewidth{0pt}
\tablehead{
\colhead{ID} &  \colhead{$N_{\rm H, Gal}^a$} & \colhead{$kT_1^b$} & \colhead{EM$_1^c$} &  \colhead{$kT_2^d$} & \colhead{EM$_2^e$} & \colhead{PL} & \colhead{$\chi^2$/dof}\\
\colhead{} &  \colhead{($10^{20}$\,cm$^{-2}$)} & \colhead{(keV)} & \colhead{($10^{-2}$\,cm$^{-6}$\,pc)}  & \colhead{(keV)} & \colhead{($10^{-2}$\,cm$^{-6}$\,pc)} &   \colhead{Norm$^f$} & \colhead{}
}
\startdata
\multicolumn{8}{c}{North bubble} \\
\tableline
N1         & 3.37 & 0.1(fix) & 5.76$\pm$1.05        & 0.304$^{+0.019}_{-0.015}$ & 6.12$\pm$0.71        & 1.02$\pm$0.06        & 189.28/155\\
N2         & 3.83 & 0.1(fix) & 5.66$\pm$1.03        & 0.320$^{+0.021}_{-0.017}$ & 5.96$\pm$0.71        & 1.01$\pm$0.07        & 171.73/155 \\
N3         & 3.86 & 0.1(fix) & 0.36$^{+6.51}_{-0.36}$ & 0.297$^{+0.029}_{-0.013}$ & 7.22$^{+0.80}_{-1.47}$ & 1.08$\pm$0.08        & 172.51/146 \\
N4         & 4.06 & 0.1(fix) & 6.78$\pm$1.10        & 0.310$^{+0.021}_{-0.017}$ & 6.17$\pm$0.76        & 0.69$\pm$0.06        & 225.82/155 \\
N5         & 4.26 & 0.1(fix) & 5.28$^{+1.07}_{-1.24}$ & 0.280$^{+0.016}_{-0.021}$ & 6.35$^{+1.24}_{-0.76}$ & 0.88$\pm$0.07        & 153.12/155 \\
N6         & 4.45 & 0.1(fix) & 7.24$\pm$1.05        & 0.304$^{+0.026}_{-0.020}$ & 4.36$\pm$0.68        & 1.01$\pm$0.06        & 169.60/155 \\
N7         & 4.76 & 0.1(fix) & 5.81$\pm$0.95        & 0.282$^{+0.018}_{-0.022}$ & 5.23$^{+1.06}_{-0.67}$ & 0.62$\pm$0.05        & 171.31/155 \\
N8         & 5.02 & 0.1(fix) & 6.05$\pm$0.93        & 0.284$\pm$0.022         & 4.28$^{+0.84}_{-0.65}$ & 0.82$\pm$0.06        & 172.76/155 \\
N\_cap\_on & 4.12 & 0.1(fix) & 3.70$\pm$0.99        & 0.307$^{+0.074}_{-0.031}$ & 2.33$^{+0.59}_{-0.71}$ & 0.96$\pm$0.07        & 187.91/149 \\
N\_cap\_off & 10.69& 0.1(fix) & 3.85$\pm$0.86        & 0.299$^{+0.025}_{-0.019}$ & 4.94$\pm$0.76        & 0.82$\pm$0.06        & 142.05/148 \\
N\_cap\_1  & 3.02 & 0.1(fix) & 1.80$^{+1.40}_{-1.39}$ & 0.245$^{+0.052}_{-0.026}$ & 2.95$^{+1.17}_{-1.08}$ & 0.81$\pm$0.06        & 191.18/150 \\
N\_cap\_2  & 4.27 & 0.1(fix) & 6.13$^{+1.90}_{-2.29}$ & 0.360$^{+0.309}_{-0.062}$ & 4.00$^{+1.11}_{-2.39}$ & 0.99$\pm$0.13        & 152.37/150 \\
N\_cap\_3  & 7.47 & 0.1(fix) & 2.28$\pm$0.97        & 0.303$^{+0.029}_{-0.022}$ & 4.31$\pm$0.75         & 0.92$\pm$0.07        & 197.40/150 \\
N\_cap\_4  & 7.82 & 0.1(fix) & 1.49$\pm$0.46        & 0.303$^{+0.017}_{-0.015}$ & 4.12$\pm$0.44         & 0.81$\pm$0.05        & 168.41/150 \\
N\_cap\_5  & 8.11 & 0.1(fix) & 2.01$\pm$0.51        & 0.289$^{+0.013}_{-0.011}$ & 5.89$\pm$0.54     & 0.77$\pm$0.06        & 161.25/149 \\
\tableline
\multicolumn{8}{c}{South bubble} \\
\tableline
S1      & 1.84 & 0.1(fix) & 4.31$^{+1.10}_{-1.47}$ & 0.283$^{+0.246}_{-0.082}$ & 0.87$^{+1.27}_{-0.54}$ & 0.90$\pm$0.07         & 156.53/142 \\
S2      & 1.66 & 0.1(fix) & 4.09$^{+1.03}_{-1.15}$ & 0.281$^{+0.111}_{-0.056}$ & 1.08$^{+0.81}_{-0.51}$ & 0.94$\pm$0.07         & 178.68/152 \\
S3      & 1.89 & 0.1(fix) & 3.63$\pm$0.57        & 0.350$\pm$0.078         & 0.90$\pm$0.30        & 0.91$\pm$0.05          & 201.77/154 \\
S4      & 2.16 & 0.1(fix) & 5.03$^{+0.86}_{-0.97}$ & 0.334$^{+0.104}_{-0.060}$ & 1.00$^{+0.49}_{-0.36}$ & 0.97$^{+0.07}_{-0.06}$  & 180.01/152 \\
S5      & 2.45 & 0.1(fix) & 4.88$^{+0.93}_{-1.07}$ & 0.256$^{+0.063}_{-0.040}$ & 1.40$^{+0.85}_{-0.55}$ & 0.86$\pm$0.05        & 188.60/155 \\
S6      & 3.03 & 0.1(fix) & 4.78$^{+1.55}_{-2.28}$ & 0.233$^{+0.107}_{-0.053}$ & 1.89$^{+2.97}_{-0.71}$ & 0.69$\pm$0.07        & 186.88/148 \\
SE\_on  & 11.87& 0.1(fix) & 9.48$\pm$1.85   & 0.300$^{+0.009}_{-0.008}$ & 28.3$\pm$1.87        & 0.65$\pm$0.08        & 192.06/150 \\
SE\_off & 11.56& 0.1(fix) & 7.00$\pm$1.19   & 0.300$^{+0.014}_{-0.012}$ & 11.6$\pm$1.09        & 0.82$\pm$0.06        & 178.33/150 \\
BULGE\_6& 10.50& 0.1(fix) & 5.12$\pm$0.90 & 0.296$^{+0.012}_{-0.011}$ & 11.5$\pm$0.97        & 0.70$\pm$0.07        & 163.92/149 \\
RXJ1856 & 9.01 & 0.1(fix) & 3.01$\pm$0.51 & 0.295$^{+0.012}_{-0.010}$ & 7.22$\pm$0.56        & 0.92$\pm$0.07        & 182.59/149 \\
EMS1274 & 5.59 & 0.1(fix) & 3.71$\pm$0.91 & 0.290$^{+0.013}_{-0.011}$ & 6.54$\pm$0.62        & 0.89$\pm$0.05        & 208.41/149 \\
EMS1388 & 5.23 & 0.1(fix) & 1.91$^{+0.99}_{-1.27}$ & 0.281$^{+0.069}_{-0.054}$ & 1.65$^{+1.16}_{-0.61}$ & 0.80$\pm$0.07        & 207.34/149 \\
RCS2118 & 2.97 & 0.1(fix) & 3.26$\pm$0.81 & 0.307$^{+0.041}_{-0.027}$ & 1.92$\pm$0.41        & 0.73$\pm$0.05        & 176.40/149 \\
NGC7130 & 2.10 & 0.1(fix) & 1.94$^{+0.60}_{-0.61}$ & 0.308$^{+0.102}_{-0.058}$ & 0.71$^{+0.37}_{-0.28}$ & 0.74$\pm$0.05        & 194.91/149 \\
\tableline
\enddata 
\tablecomments{
$^a$: The absorption column densities for the CXB and the GH/NPS components (\textsc{wabs*(apec2 + pl)}) were fixed to Galactic values given in Dickey \& Lockman (1990).\\
$^b$: Temperature of the LB/SWCX plasma fitted with the \textsc{apec} model for the fixed abundance $Z = Z_{\odot}$.\\
$^c$: Emission measure of  the LB/SWCX  plasma fitted with the \textsc{apec} model for the fixed abundance $Z = Z_{\odot}$.\\
$^d$: Temperature of the GH/NPS plasma fitted with the \textsc{apec} model for the fixed abundance $Z = 0.2 \, Z_{\odot}$.\\
$^e$: Emission measure of the GH/NPS plasma fitted with the \textsc{apec} model for the fixed abundance $Z = 0.2 \, Z_{\odot}$.\\
$^f$: The normalization of the PL in units of $5.85 \times 10^{-8}$\,erg\,cm$^{-2}$\,s$^{-1}$\,sr$^{-1}$, given in Kushino et al. (2012) as an average of 91 observation fields, assuming a single power-law model with a photon index $\Gamma_{\rm CXB} = 1.41$.}
\end{deluxetable*}


\begin{deluxetable*}{lccccccc}
\tabletypesize{\scriptsize}
\tablecaption{Fitting parameters for \SW observations}
\tablewidth{0pt}
\tablehead{
\colhead{ID} &  \colhead{$N_{\rm H, Gal}^a$} & \colhead{$kT_1^b$} & \colhead{EM$_1^c$} &  \colhead{$kT_2^d$} & \colhead{EM$_2^e$} & \colhead{PL} & \colhead{$\chi^2$/dof}\\
\colhead{} &  \colhead{($10^{20}$\,cm$^{-2}$)} & \colhead{(keV)} & \colhead{($10^{-2}$\,cm$^{-6}$\,pc)}  & \colhead{(keV)} & \colhead{($10^{-2}$\,cm$^{-6}$\,pc)} &   \colhead{Norm$^f$} & \colhead{}
}
\startdata
\multicolumn{8}{c}{North bubble} \\
\tableline
50\deg $<$ $b$ $<$ 55\deg & 3.67 & 0.1(fix) & 2.87$^{+0.51}_{-0.54}$ & 0.327$^{+0.067}_{-0.037}$ & 2.05$^{+0.48}_{-0.47}$ & 1.73$^{+0.12}_{-0.13}$ & 41.24/27 \\
45\deg $<$ $b$ $<$ 50\deg & 4.71 & 0.1(fix) & 4.24$^{+1.11}_{-1.19}$ & 0.273$^{+0.051}_{-0.033}$ & 3.60$^{+1.34}_{-1.07}$ & 2.25$^{+0.18}_{-0.19}$ & 58.97/39 \\ 
40\deg $<$ $b$ $<$ 45\deg & 4.71 & 0.1(fix) & 2.08$^{+0.73}_{-0.76}$ & 0.294$^{+0.052}_{-0.038}$ & 2.86$^{+0.84}_{-0.73}$ & 2.03$\pm$0.13         & 54.01/39 \\
35\deg $<$ $b$ $<$ 40\deg & 7.84 & 0.1(fix) & 2.54$^{+0.52}_{-0.60}$ & 0.273$\pm$0.023 & 4.24$^{+0.93}_{-0.65}$ & 1.96$\pm$0.10         & 58.68/39 \\
20\deg $<$ $b$ $<$ 35\deg & 11.16 & 0.1(fix) & 2.64$\pm$0.47 & 0.294$^{+0.023}_{-0.018}$ & 5.82$\pm$0.76 & 1.60$\pm$0.14         & 40.23/27 \\
15\deg $<$ $b$ $<$ 20\deg & 12.83 & 0.1(fix) & 1.25$^{+0.62}_{-0.66}$ & 0.277$^{+0.026}_{-0.023}$ & 7.11$^{+1.49}_{-1.19}$ & 2.01$\pm$0.14         & 58.15/39 \\
10\deg $<$ $b$ $<$ 15\deg & 14.79 & 0.1(fix) & 1.20$\pm$0.44 & 0.315$^{+0.022}_{-0.018}$ & 8.65$\pm$0.99        & 1.81$\pm$0.17         & 53.35/26 \\
5\deg $<$ $b$ $<$ 10\deg  & 24.74 & 0.1(fix) & 1.86$\pm$0.56 & 0.287$^{+0.026}_{-0.021}$ & 14.6$^{+2.60}_{-2.41}$ & 2.73$\pm$0.18         & 44.93/39 \\
Swift16 (NPS)   & 4.50 & 0.1(fix) & 5.46$\pm$1.99 & 0.303$^{+0.053}_{-0.036}$ & 7.38$^{+1.99}_{-1.92}$ & 1.82$\pm$0.24         & 44.01/39 \\
Swift19 (NPS)      & 5.70 & 0.1(fix) & 4.08$\pm$1.74 & 0.291$^{+0.028}_{-0.023}$ & 12.1$\pm$2.00        & 2.03$\pm$0.22         & 50.58/39 \\
AS210 (NW-clump)     & 15.79 & 0.1(fix) & 2.73$\pm$1.01 & 0.294$^{+0.020}_{-0.016}$ & 20.4$\pm$2.51 & 1.94$\pm$0.19         & 63.09/39 \\
IGRJ1648 (NW-clump)  & 17.56 & 0.1(fix) & 1.91$\pm$1.12 & 0.299$^{+0.026}_{-0.020}$ & 21.7$\pm$3.21        & 2.73$\pm$0.26         & 54.79/39 \\
\tableline
\multicolumn{8}{c}{South bubble} \\
\tableline
$-$15\deg $<$ $b$ $<$ $-$10\deg & 13.51 & 0.1(fix) & 3.81$\pm$0.76 & 0.312$^{+0.019}_{-0.015}$ & 14.8$\pm$1.52        & 2.21$\pm$0.13        & 85.64/39 \\
$-$20\deg $<$ $b$ $<$ $-$15\deg & 8.66 & 0.1(fix) & 3.54$^{+1.21}_{-1.25}$ & 0.289$^{+0.034}_{-0.026}$ & 8.29$^{+1.88}_{-1.66}$ & 2.03$\pm$0.19        & 37.91/39 \\
$-$25\deg $<$ $b$ $<$ $-$20\deg & 6.77 & 0.1(fix) & 1.67$^{+0.76}_{-0.82}$ & 0.273$^{+0.016}_{-0.016}$ & 7.80$^{+1.16}_{-0.92}$ & 1.98$\pm$0.13        & 49.11/39 \\
$-$35\deg $<$ $b$ $<$ $-$25\deg & 5.41 & 0.1(fix) & 2.21$^{+0.50}_{-0.83}$ & 0.268$^{+0.029}_{-0.038}$ & 2.52$^{+1.10}_{-0.52}$ & 2.01$\pm$0.10        & 64.90/39 \\
$-$45\deg $<$ $b$ $<$ $-$35\deg & 5.16 & 0.1(fix) & 2.35$^{+0.46}_{-0.78}$ & 0.267$^{+0.026}_{-0.038}$ & 2.29$^{+1.00}_{-0.46}$ & 2.14$^{+0.10}_{-0.09}$ & 62.04/39 \\
$-$50\deg $<$ $b$ $<$ $-$45\deg & 3.04 & 0.1(fix) & 1.63$^{+0.60}_{-0.78}$ & 0.247$^{+0.052}_{-0.042}$ & 1.65$^{+0.94}_{-0.59}$ & 1.80$\pm$0.09        & 57.65/39 \\
$-$55\deg $<$ $b$ $<$ $-$50\deg & 1.56 & 0.1(fix) & 1.87$^{+0.68}_{-1.17}$ & 0.233$^{+0.110}_{-0.059}$ & 0.89$^{+1.45}_{-0.55}$ & 1.70$^{+0.10}_{-0.11}$ & 51.25/39 \\
PBCJ1847 (SE-claw)    & 14.71 & 0.1(fix) & 2.38$^{+1.78}_{-1.79}$ & 0.323$^{+0.030}_{-0.025}$ & 24.6$^{+4.15}_{-3.78}$ & 1.72$\pm$0.30        & 46.53/39 \\
PBCJ1919 (SE-claw)    & 9.13  & 0.1(fix) & 2.57$\pm$1.79  & 0.306$^{+0.042}_{-0.029}$ & 12.9$^{+2.75}_{-2.72}$ & 2.21$\pm$0.28        & 50.78/39 \\
\tableline
\enddata 
\tablecomments{
$^a$: The absorption column densities for the CXB and the GH/NPS components (\textsc{wabs*(apec2 + pl)}) were fixed to Galactic values given in Dickey \& Lockman (1990).\\
$^b$: Temperature of the LB/SWCX plasma fitted with the \textsc{apec} model for the fixed abundance $Z = Z_{\odot}$.\\
$^c$: Emission measure of  the LB/SWCX  plasma fitted with the \textsc{apec} model for the fixed abundance $Z = Z_{\odot}$.\\
$^d$: Temperature of the GH/NPS plasma fitted with the \textsc{apec} model for the fixed abundance $Z = 0.2 \, Z_{\odot}$.\\
$^e$: Emission measure of the GH/NPS plasma fitted with the \textsc{apec} model for the fixed abundance $Z = 0.2 \, Z_{\odot}$.\\
$^f$: The normalization of the CXB in units of $5.85 \times 10^{-8}$\,erg\,cm$^{-2}$\,s$^{-1}$\,sr$^{-1}$, given in Kushino et al. (2012) as an average of 91 observation fields, assuming a single power-law model with a photon index $\Gamma_{\rm CXB} = 1.41$.}
\end{deluxetable*}

{}

\end{document}